\documentclass[aps, pre, twocolumn, showpacs]{revtex4}
\usepackage{graphicx}
\usepackage{amsmath}
\usepackage{amssymb}
\usepackage{wasysym}

\def\e{\ensuremath{\mathrm{e}}}
\def\nn{\nonumber}
\def\ba{\begin{eqnarray}}
\def\ea{\end{eqnarray}}
\def\Fix{\mathrm{Fix}}
\def\S{\ensuremath{\mathcal{S}}}

\begin{document}
 
\title{The meandering instability of a viscous thread}

\author{Stephen W. Morris$^{1}$, Jonathan H. P. Dawes$^{2}$, Neil M. Ribe$^{3}$,  and John R. Lister$^{2}$  }
%
% Is this author order still reasonable?
%
%
%%%  LaTeX warns:
% Class revtex4 Warning: Command \address is obsolete;
% Use \affiliation instead.
%
\affiliation{$^{1}$Department of Physics, University of
Toronto, 60 St. George St., Toronto, Ontario, Canada M5S 1A7\\
$^{2}$ DAMTP, Centre for Mathematical Sciences, University of Cambridge, Wilberforce Road, Cambridge CB3 0WA  \\
$^{3}$ Dynamique des Fluides G\'eologiques, IPGP et Universit\'e de Paris-7, CNRS,
Tour 14, 2, place Jussieu, 75005 Paris, France
}

\date{\today}

\begin{abstract}
%
%  This is probably rather long for PRE
%
A viscous thread falling 
from a nozzle onto a surface exhibits the famous rope-coiling effect, in which the thread buckles to form loops.  If the surface is replaced by a belt moving with speed $U$, the rotational symmetry of the buckling instability is broken and a wealth of interesting states are observed [See S. Chiu-Webster and J. R. Lister, J. Fluid Mech., {\bf 569}, 89 (2006)].
% Moved reference to the relevant sentence JRL
We experimentally studied this ``fluid mechanical sewing machine''  in a new, more precise apparatus.
As $U$ is reduced, the 
steady catenary thread bifurcates into a meandering state in which the thread displacements are only transverse to the motion of the belt.  We measured the amplitude and frequency $\omega$  of the meandering close to the bifurcation.  For smaller $U$, single-frequency meandering bifurcates to a two-frequency ``figure eight'' state, which contains a significant $2\omega$ component and parallel as well as transverse displacements. This eventually reverts to single-frequency coiling at still smaller $U$.  
More complex, highly hysteretic states with additional frequencies are observed for larger nozzle heights. 
We propose to understand this zoology in terms of the generic amplitude equations appropriate for resonant interactions between two oscillatory modes with frequencies $\omega$ and $2\omega$.
% and $3\omega$. 
 % Reference to 3omega omitted - we don't include this in the amp eqns
 % later. OK - JHPD.   
The form of the amplitude equations captures both the axisymmetry of the $U=0$ coiling state and the symmetry-breaking effects induced by the moving belt.
\end{abstract}

% 82.40.Bj     Oscillations, chaos, and bifurcations 
% 05.45.-a     Nonlinear dynamics and chaos  
% 47.52.+j     Chaos in fluid dynamics 
% 47.20.Gv   Viscous and viscoelastic instabilities 
% 47.20.-k     Flow instabilities 
% 47.20.Ky    Nonlinearity, bifurcation, and symmetry breaking
% 47.54.-r      Pattern selection; pattern formation 
% Choice of PACS seems fine to me! JHPD.

\maketitle

The fascinating instabilities of very viscous fluids are familiar to anyone who has  poured syrup onto a pancake~\cite{jay_ingram}.  The thread of syrup elongates as it falls, and exhibits a buckling instability as nears the surface~\cite{taylor}.  The buckling  is the result of a competition between bending and axial compression, and causes the thread to loop itself into coils.  This ``liquid rope coiling effect''~\cite{barnes_woodcock} has a long history comprising almost a half century of experiments~\cite{taylor,barnes_woodcock,barnes_mckenzie,cruickshank_munson, maleki_PRL, mult_states}, scaling analyses~\cite{maha, maha_corr, mult_states}, analytic theory and numerical simulation~\cite{mult_states,cruickshank, tchavdarov, maha_EPL, ribe_PRSL, coiling_stability}.  It has been observed that the coils tend to pile up on the surface, forming a tall column~\cite{barnes_woodcock,barnes_mckenzie, maleki_PRL} and that the frequency of coiling exhibits a complex, multivalued dependence on the fall height~\cite{maleki_PRL,mult_states}.  Here we consider a variation on this classic experiment, the case when the liquid thread falls onto a moving surface, a situation known as the ``fluid-mechanical sewing machine''~\cite{chu_webster}
owing to the variety of patterns deposited on the surface.
%  added explanatory phrase JRL

The moving surface breaks one of the basic symmetries of the ``pure'' rope coiling problem, and has the effect of unfolding the coiling instability into a rich panoply of distinct bifurcations. This arrangement is also very convenient experimentally; the moving surface is provided by a belt which is cleared of fluid before it returns, eliminating the pile-up of viscous fluid at the point of contact.

 The main experimental control parameters are the height $H$ of the
 nozzle from which the thread descends, and the speed of the moving
 surface $U$.  Less important parameters, which we hold
%
% 'held' changed to 'hold' to preserve use of present tense throughout
% this paragraph - JHPD. 
%
fixed in this
 work, are the volumetric flow rate of the liquid $Q$ and the diameter
 of the nozzle $d$.  The Newtonian fluid is characterized by its
 density $\rho$, kinematic viscosity $\nu$ and surface tension
 $\sigma$.  Non-Newtonian effects are negligible.

We present an experimental study of the ``stitch'' patterns made by
the thread as it is laid down on the belt, as a function of $H$ and
$U$.  We first discuss a survey of the states in the $H-U$ plane, with
denser coverage than was possible in Ref.~\cite{chu_webster}.  Then we
focus on the simplest bifurcation, that from the straight catenary
state to meandering.  We show that this state is well described as a
forward Hopf bifurcation and compare the onset belt speed and Hopf
frequency to recent linear stability
theory~\cite{meander_linear_stability}.   We examine the nonlinear
saturation of the meandering amplitude and qualitatively explain it
with a simple kinematic model of an inextensible thread.  We then
essay the task of understanding the more complex ``figure 8'' and other
patterns that appear as the belt speed is lowered.  We propose a
general amplitude equation approach, based on the idea that the motion
of the belt can be treated as a perturbation to the pure rope coiling problem which has $O(2)$ symmetry.
%treated as a perturbation on the pure rope
%coiling effect which has $O(2)$ symmetry.  
We propose a set of coupled
amplitude equations which minimally break $O(2)$ symmetry, and mix
amplitudes for rotating wave modes with frequencies $\omega$ and $2
\omega$.
%
% $O(2)$ should always be set in a math environment - JHPD.
%
We conjecture that all the complicated stitch patterns can
be captured by this approach, which could be generalized to include
more modes as they become excited near the multifrequency
regime. These modes can be directly measured by visualizing
longitudinal and transverse motions of the thread by viewing from the
side.

This paper is organized in the following way.  In section \ref{expt},
we discuss the fluid properties and the experimental apparatus.  In
section \ref{results}, we make comparisons between our parameters and
the corresponding zero belt speed, pure coiling problem, before
turning to the state diagram in section \ref{states_section}, and the
meandering threshold in sections \ref{sec:meanderingonset} and
\ref{amp_model}.  In section \ref{beyond}, we discuss experimental
observations of the states
beyond meandering. Section~\ref{sec:ampeqns} presents our general
amplitude formalism which provides a consistent framework for
interpreting the experimental results. We have placed a
detailed discussion of the derivation of the amplitude equations 
in an Appendix. Section \ref{conc} is a brief conclusion.

\section{experiment}
\label{expt}
\begin{figure}
\includegraphics[height=6cm]{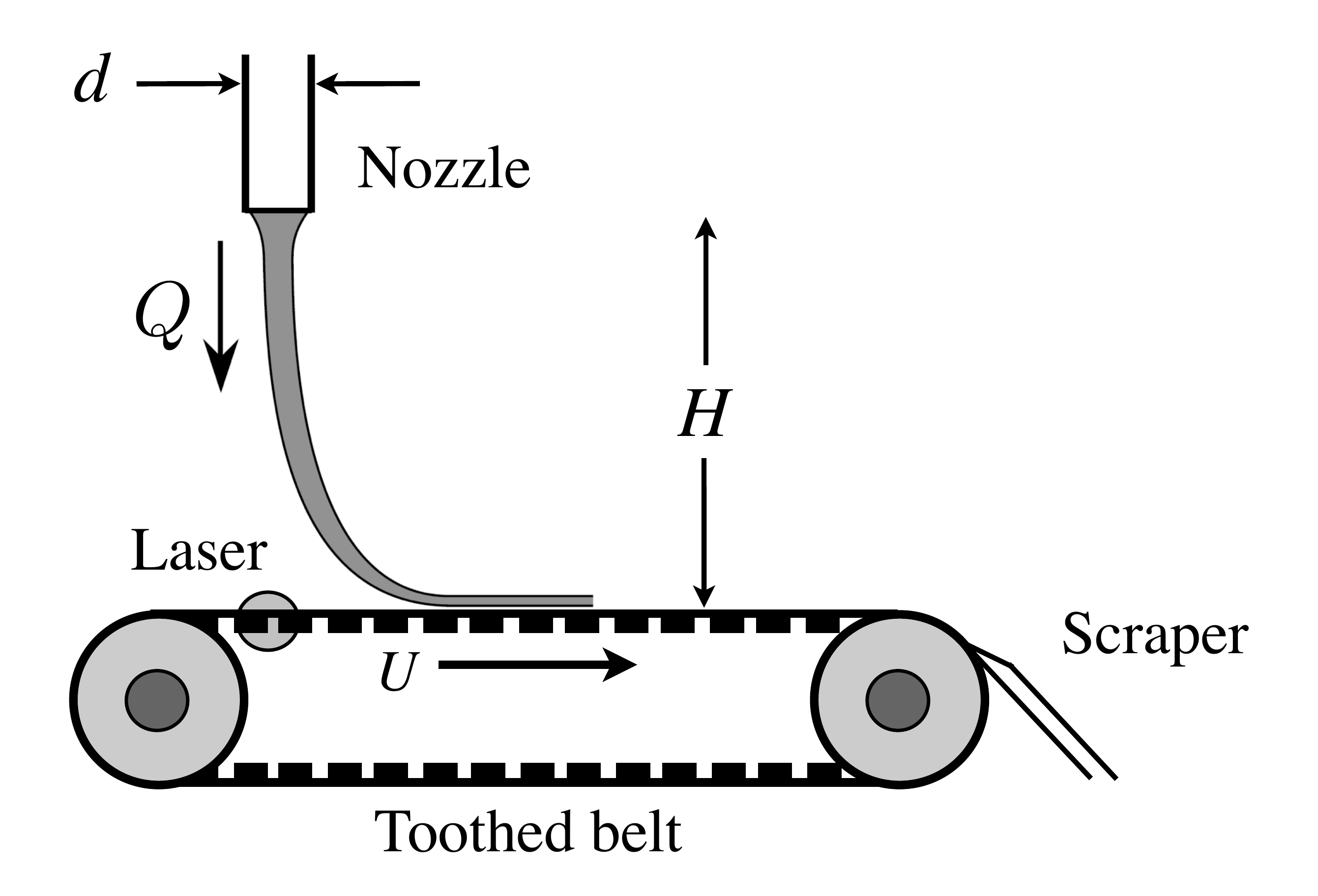}
\caption{A schematic of the apparatus.  A silicone oil stream with volumetric flow rate $Q$ falls a distance $H$ from a nozzle of diameter $d$ onto a belt moving at speed $U$.  The oil is removed by the scraper.  The laser, which was interrupted by the teeth of the belt, was used to measure the speed $U$.
}
\label{apparatus}
\end{figure}

Our experimental apparatus was a redesigned version of the one described in Ref.~\cite{chu_webster}.  It is shown schematically in Fig.~\ref{apparatus}.  The main differences were improved control and measurement of the belt speed, and the use of very stable
silicone oil as the working fluid.

Silicone oil offers several advantage over the Lyle's Golden Syrup used in previous experiments~\cite{chu_webster}.  Although its viscosity cannot be varied by adding water as described in Ref.~\cite{chu_webster}, silicone oil is not susceptible to evaporation and 
its viscosity varies much less with temperature than that of sugar syrup.  This makes it possible to do highly reproducible experiments over a long period of time. It also has about one third the surface tension of syrup, which reduces the thread-thread and thread-belt interaction after the fluid is laid down on the belt.

We used Dow  Corning 200\copyright ~fluid, which has a nominal kinematic viscosity of 30,000~cSt.  To fully characterize the fluid, we measured its viscosity and the temperature coefficient of its viscosity using a TA AR1000 rheometer.  The same instrument was used to confirm that the fluid is Newtonian to an excellent approximation.  We also measured the fluid density and its temperature coefficient using an Anton-Parr DMA 5000 densitometer.  The results of these measurements are shown in Table~\ref{si_oil_props}.  

The temperature of the experiment was recorded during each run, but not otherwise controlled.  The only significant variation in the fluid properties was therefore due to the temperature, and not to the degradation of the fluid.  Averaged over all experiments, the mean temperature was $22.2 \pm 0.8^\circ$C.  Accounting for this systematic error, the fluid had a density of $\rho = 996 \pm 8$~kg/m$^3$ and a kinematic viscosity of $\nu = (2.77 \pm 0.02) \times 10^{-2}$~m$^2$/s, or $27,700 \pm 200$ cSt.

\begin{table}
\begin{tabular}{|l|c|}
\hline
Density $\rho$  							&  $996 \pm 8$~kg/m$^3$ \\ % adjusted to expt T
Temperature coefficient of $\rho$ 			&  $-0.0885~\%/^\circ$C \\
Molecular viscosity $\eta$ 				& $27.63 \pm 0.05$ Pa~s \\ % adjusted to expt T
Temperature coefficient of $\eta$ 			& $-1.6~\%/^\circ$C \\
Kinematic viscosity $\nu = \eta/\rho$			& $(2.77 \pm 0.02) \times 10^{-2}$~m$^2$/s \\
\hline
Maxwell time $\tau_M$						& $< 1.8$ ms \\
Young's modulus $G$					& $< 14$ kPa \\
\hline 
Surface tension $\sigma$ 				 	&  21.5~mN/m \\
\hline
\end{tabular}
%\tablenotetext{a}{From Dow Corning data sheet.}
\caption{\label{si_oil_props} The measured properties of the silicone oil at the mean experimental temperature of $22.2\pm0.8^\circ$~C.  The size of the measured Maxwell time and Young's modulus show that viscoelastic effects are negligible.  The surface tension is taken from the manufacturer's data sheet.}
%
% data from Lucas' notes:
%The viscosity of silicone oil at 26.4 C is 27.56 ±0.05 Pa.s.
% The temperature dependence is 1.61 % / degree.
% 
%Temperature (C)	density (g/ml)	residuals (ug/ml)	residuals (ug/ml)	
%20	0.975579	-67	6.836	-1.62362452
%22	0.973784	-41.8	6.39496	0.5292427669
%24	0.971994	-21.6	4.54944	0.976877767
%26	0.970209	-6.4	1.29944	-0.2867744112
%28	0.968428	4.8	-2.35504	-2.267757531
%30	0.966647	16	-2.414	-0.9721042507
%32	0.964868	25.2	-0.87744	1.59416386
%34	0.963097	26.4	-3.74536	-0.5749637038
%36	0.961327	26.6	-4.01776	-0.4854864012
%38	0.95956	23.8	-3.69464	-0.1433926659
%40	0.957796	18	-2.776	0.4453400735
%42	0.956037	7.2	-3.26184	-0.7252546267
%44	0.954281	-6.6	-3.15216	-1.661132245
%46	0.952524	-19.4	1.55304	1.631762684
%48	0.950774	-39.2	2.85376	1.14749655
%50	0.949025	-60	6.75	2.880146648 
%\begin{tabular}{|l|l|}
%
%  Parameters analysis in matlab file called T_rho_rQ_eta.m
%
%
\end{table}

 We used an internally reinforced, toothed timing belt of width 16~mm.  The belt was looped over toothed pulleys and driven with a speed-controlled DC motor.  The speed of the belt $U$ was measured using an averaging storage oscilloscope to record the intensity of a laser beam that was interrupted by the teeth of the belt. The frequency of the passage of the teeth, measured by an internal function of the oscilloscope, could be used to find $U$ to within a few \%.  The reinforced timing belt resisted stretching or slipping, even under considerable pressure from the plastic scraper used to remove the silicone oil.  After scraping, the belt had only a very thin wetting layer of oil left on it.  The excess oil dripped off the scraper into a beaker, and could be recirculated.
 
In all the experiments reported here, we used a 
nozzle with $d=8.00\pm0.02$~mm  to deliver a fixed, constant volumetric flux of oil $Q$.  The oil was supplied by a reservoir and driven by a gravity feed system with a fixed 42~cm head.  The reservoir and the nozzle were attached to a frame mounted on a screw so that they travelled together, maintaining a constant head as the fall height $H$ was varied.  $H$ was measured with a scale to $\pm 0.05$~cm.

Before and after each run, we weighed the quantity of oil delivered in a certain time interval, typically 2~min, timed with a stopwatch.  We found the mass flux $\rho Q = 0.0270 \pm 0.0006$~g/s remained constant within experimental error, even for runs performed many days apart.  All the experiments we discuss here used this value of $\rho Q$.

\begin{figure}
\includegraphics[width=8cm]{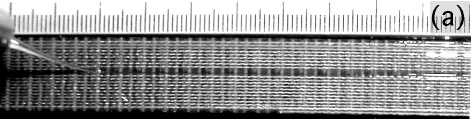}
\includegraphics[width=8cm]{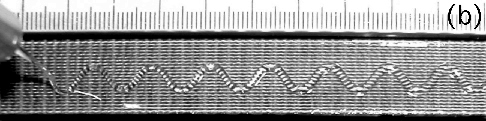}
\includegraphics[width=8cm]{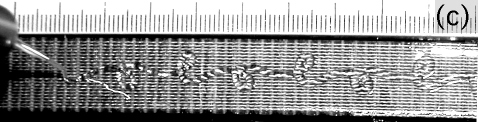}
\includegraphics[width=8cm]{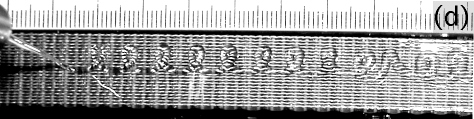}
\caption{Photos of some of the states of the thread on the belt.  The nozzle is on the left and the belt moves from left to right.  The internal reinforcing mesh of the semi-transparent belt is visible, but the surface of the belt is smooth.  The scale is in mm. (a) The catenary steady state. (b) The meandering state. (c) The figure eight state. (d) The translated coiling state. }
\label{state_photos}
\end{figure}

An experimental run consisted of fixing the nozzle height $H$ and varying the belt speed $U$, starting from high values of $U$ for which the thread is dragged horizontally by the belt to form a steady, stretched catenary.  Lowering the belt speed to zero, and then increasing it again produces a sequence of bifurcations, and a sequence of ``stitch patterns'' of the thread on the belt.  Care was taken to change $U$ slowly near the onset of the first instability, to meandering, where 
the state of the thread responded very slowly to small changes in $U$.
We recorded the oil pattern on the moving belt from above using a 3 megapixel digital camera.  To avoid heating the oil, and to freeze the motion of the thread, the only lighting used was a fast photo flash.  Typical photos of
the main states of interest are shown in Fig.~\ref{state_photos}.  These and many other states are shown  and described in Ref.~\cite{chu_webster}.

%% two quotes here simply to get my EMACS version
%% to continue colouring plain text black. '' JHPD.

\section{results and discussion}
\label{results}

In this section we will tour the zoology of states and bifurcations exhibited by the viscous thread falling on the moving belt.  After a general overview, we focus on the simplest transition first, the onset of the meandering state from the stretched catenary as the belt speed is lowered.  We compare our results to the existing linear stability theory for this transition~\cite{meander_linear_stability}.  Moving into the nonlinear regime, we find that the amplitude of the meandering is well described as a forward Hopf bifurcation.

We then show that the saturation coefficient of the amplitude can be
deduced reasonably well from a simple model of an inextensible thread.
Finally, we consider the subsequent bifurcations for small nozzle
heights $H$, namely the appearance of the figure eight and translated
coiling states.  These and the other more intricate states for larger
$H$ are probably best understood from the dynamics of their
% complex % - removed, JHPD.
amplitudes.  We present a brief outline of a bifurcation theoretic approach to this problem.

At the outset, it is instructive to identify the states of
``pure'' rope coiling that we would expect to see
if the speed of the belt in our experiments were zero.  Theory and 
laboratory experiments show~\cite{ribe_PRSL,mult_states,coiling_stability}, 
that pure coiling can occur in four different regimes 
- viscous (V), gravitational (G), inertio-gravitational (IG),
and inertial (I) - depending on the relative magnitudes of the different
forces acting on the thread.
%rope. 
In the V, G, and I regimes, the coiling frequency $\Omega$ is
a single-valued function of the fall height $H$. In IG coiling, by contrast, 
$\Omega (H)$ is multivalued, reflecting the fact that the nearly vertical upper part of 
the thread
%rope 
acts as a distributed pendulum with multiple resonantly excited eigenmodes.

 In a typical experiment where
the fluid properties, the flow rate $Q$ and the nozzle diameter $d$ are held constant, 
the regime that obtains is determined by the fall height $H$. As $H$ increases,
the regimes always succeed each other in the order V-G-IG-I. However, while the 
G and I regimes can be observed for nearly any choice of 
$Q$, $d$, and the working fluid, the V and IG regimes 
are confined to more restricted regions of parameter space.
For a given experiment, this space  is completely characterized by the 
three dimensionless groups,
\begin{eqnarray}
\Pi_1 &=& \bigg(\frac{\nu^5}{g Q^3}\bigg)^{1/5}  \;\; [610 \pm 10] \label{pi_1}\\ 
\Pi_2 &=& \bigg(\frac{\nu Q}{g d^4}\bigg)^{1/4}  \;\;\; [0.370 \pm 0.002] \label{pi_2}\\
\Pi_3 &=&  \frac{\sigma d^2}{\rho\nu Q}  \;\;\;\;\;\;\;\;\;\;\; [1.84 \pm 0.06]
%\Pi_1 &=& \bigg(\frac{\nu^5}{g Q^3}\bigg)^{1/5}  \;\; [610 \pm 10] \label{pi_1}\\ 
%\Pi_2 &=& \bigg(\frac{\nu Q}{g d^4}\bigg)^{1/4}  \;\;\; [0.37 \pm 0.0025] \label{pi_2}\\
%\Pi_3 &=&  \frac{\sigma d^2}{\rho\nu Q}  \;\;\;\;\;\;\;\;\;\;\; [1.8 \pm 0.044]
%
%  I checked the error bars and sig figs on these again.  Found an error in my previous numbers ...
%  SM
%
\label{pi_3}
\end{eqnarray}
where $g$ is the acceleration due to gravity, $\sigma$ is the surface tension, and the
values of each group for our experiments are given in square brackets. For a fluid
such as silicone oil with a relatively low surface tension, the value of $\Pi_3$ has
only a minor effect on the coiling behavior, which is controlled primarily by 
$\Pi_1$ and $\Pi_2$. It turns out that  our values
$\Pi_1\approx 610$ and $\Pi_2\approx 0.37$ are within the domain where IG
coiling is expected to occur for some range of heights (See Ref.~\cite{mult_states}, Fig.~3.)
By contrast, the V regime does not occur, because the characteristic height $(\nu Q/g)^{1/4}\approx$ 3~mm over which gravitational stretching of the thread becomes important is less than the diameter of the nozzle 
(as usual with a gravity feed system).
% Poiseuille flow gives (\nu Q/g)^{1/4}=0.8a and even having a large reservoir 
% and short nozzle doesn't get you far with a 1/4 power. JRL
A direct numerical calculation using the method
of \cite{ribe_PRSL} shows that G coiling is expected in our experiments for 2~cm $< H<$ 5.7~cm, 
IG coiling for 5.7~cm $< H<$ 8.5~cm, and I coiling for $H>$ 8.5 cm. The
experiments reported here explore the range 2~cm $< H<$ 8~cm, with
belt speeds up to $U = 10$~cm/s.

% the following equation is no longer needed here, but feel free to place it elsewhere
% if necessary.
%
%  Just omit it.  SM
%
%\begin{equation}
%h = H(g/\nu^2)^{1/3} = H~/~ (4.3~ {\rm cm}),  % check this.  Add error bar?
%\end{equation}    

%
%
% post-referee  : make the diagram wider and suitable for two columns
%
%

\begin{widetext}

\begin{figure}
\includegraphics[width=15.5cm]{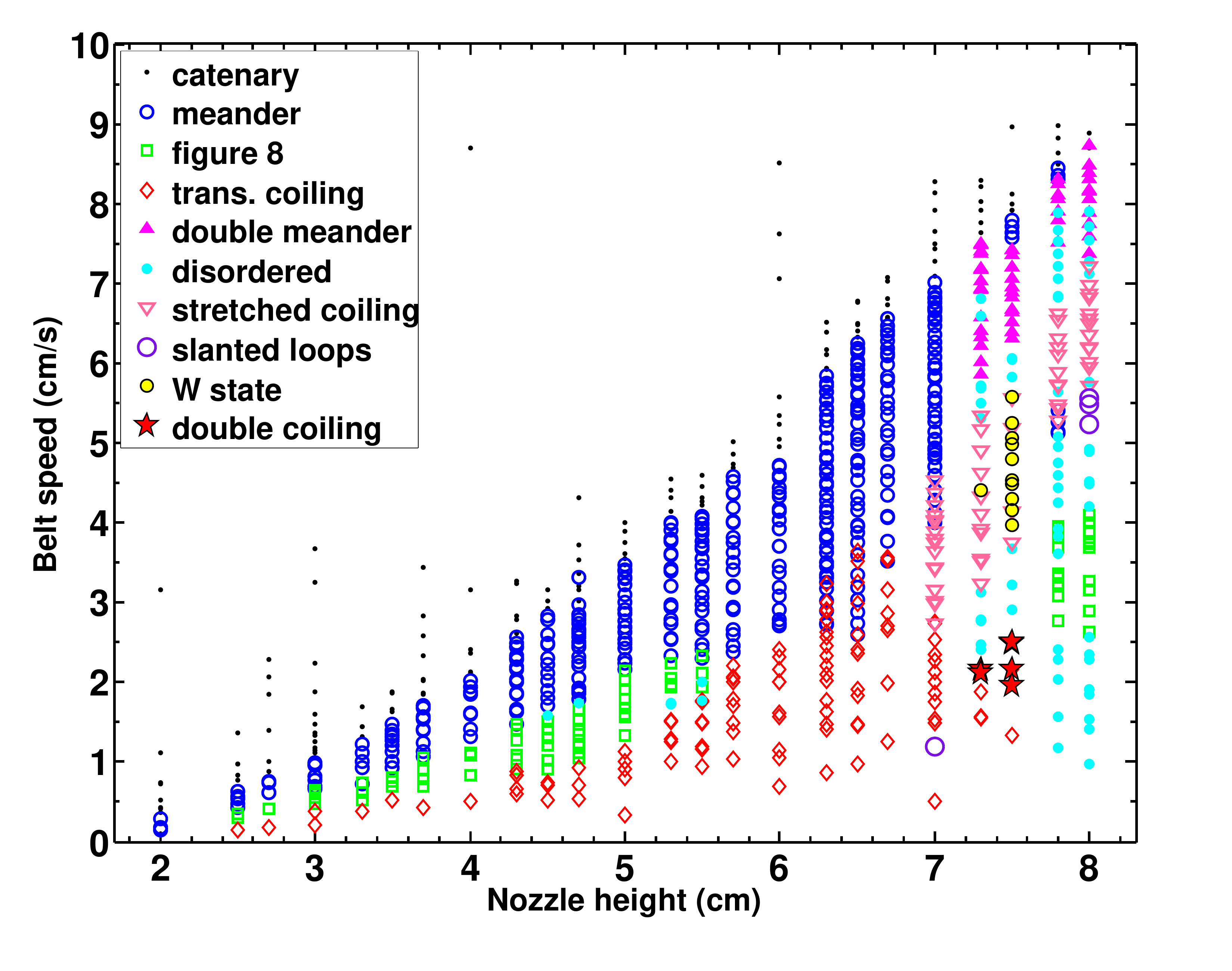}
\caption{The state diagram, as a function of the belt speed $U$ and the nozzle height $H$. Some of the simpler states are shown in the previous figure; more complex states are described in Ref.~\cite{chu_webster}.
}
\label{states}
\end{figure}

\end{widetext}

\subsection{The state diagram} 
\label{states_section}

Fig.~\ref{states} shows the states observed in the $U$-$H$ plane.  The diagram was made by scrutinizing over 1100 individual photos.  In general, the diagram agrees very well with Fig.~6 of Ref.~\cite{chu_webster}, although certain details differ at larger $H$.  At large $U$, only the stretched catenary steady state is seen.  As $U$ is reduced at fixed $H$, the first bifurcation is to the meandering state.  This transition, which is non-hysteretic for $H \leq 7.0$~cm, is discussed in detail below.  Above $H=7.0$~cm, the meandering state shows clear multifrequency behavior and a small hysteresis at onset.  For $H \le 5.5$~cm, as $U$ is reduced, the meandering state switches to a figure eight mode, before making a second transition to translated coiling. All these transitions are very nearly non-hysteretic. It is not clear if the figure eight window extends all the way down to the onset of any instability, just below $H = 2.0$~cm.  For $5.7~{\rm cm} \leq H \leq 7.0~{\rm cm}$, the figure eight state disappears, and there is a small hysteretic region where meandering and coiling co-exist. For $H \ge 7.0$~cm, many complex states are observed, with windows of various patterns separated by disordered, possibly chaotic regions.  In contrast to Ref.~\cite{chu_webster}, almost no region of slanted loops is observed.  This may be due to the reduced surface tension of the silicone oil, which reduces the tendency of the thread to stick to itself. 
%
%
%  We need to add something about bunched-up meandering --- which I did not see much.
%   it comes up in the theory later.  SM
%
%
Regions of W-states and double coiling are seen, as well as small re-entrant regions of the simpler figure eight and single-frequency meandering modes.  The latter was not observed in Ref.~\cite{chu_webster}.   We do not observe side-kicks, which may require still larger $H$. 

A complete elucidation of the multifrequency states for $H \ge 7.0$~cm must await more precise and better instrumented experiments.  For the remainder of this paper, we will focus on the simpler bifurcations between the steady and meandering states, and on the region of the  figure eight window for  $H \le 5.5$~cm.

\subsection{The threshold of single frequency meandering}

\label{sec:meanderingonset}

We now consider the bifurcation from the steady stretched state to
meandering,  which has been analyzed at the linear stability level in
Ref.~\cite{meander_linear_stability}.  In particular, linear theory
provides clear predictions for the critical belt speed $U_c$ at which
meandering begins as $U$ is decreased, and for the critical frequency
$\omega_c$ of the meandering state which emerges.    Quite remarkably,
the onset angular frequency  $\omega_c$ (which, as we will show, is
the Hopf frequency that obtains for a {\it zero} amplitude of
meandering) is predicted to be nearly identical to the frequency
expected for pure, {\it nonzero} amplitude rope coiling onto a
stationary surface for the same
$H$~\cite{meander_linear_stability}.

To accurately locate the onset of meandering, we photographed the state while very slowly decreasing and increasing $U$ through $U_c$.  We then found the peak-to-peak amplitude $2|A|$
% 
% SWM: I put lambda and 2A back here: see later
% Changed A to |A| to agree better with amp eqns here and later on. JHPD.
%
and wavelength  $\lambda$  of the meandering thread directly from the digital photos.  It was typically possible to use data spanning 5 to 10 wavelengths, and the resolution of the images was sufficiently high that the width of the thread was approximately 10 pixels. The smallest amplitudes we could measure were a fraction of the width of the thread, while the largest spanned the belt.

\begin{figure}
\includegraphics[width=8cm]{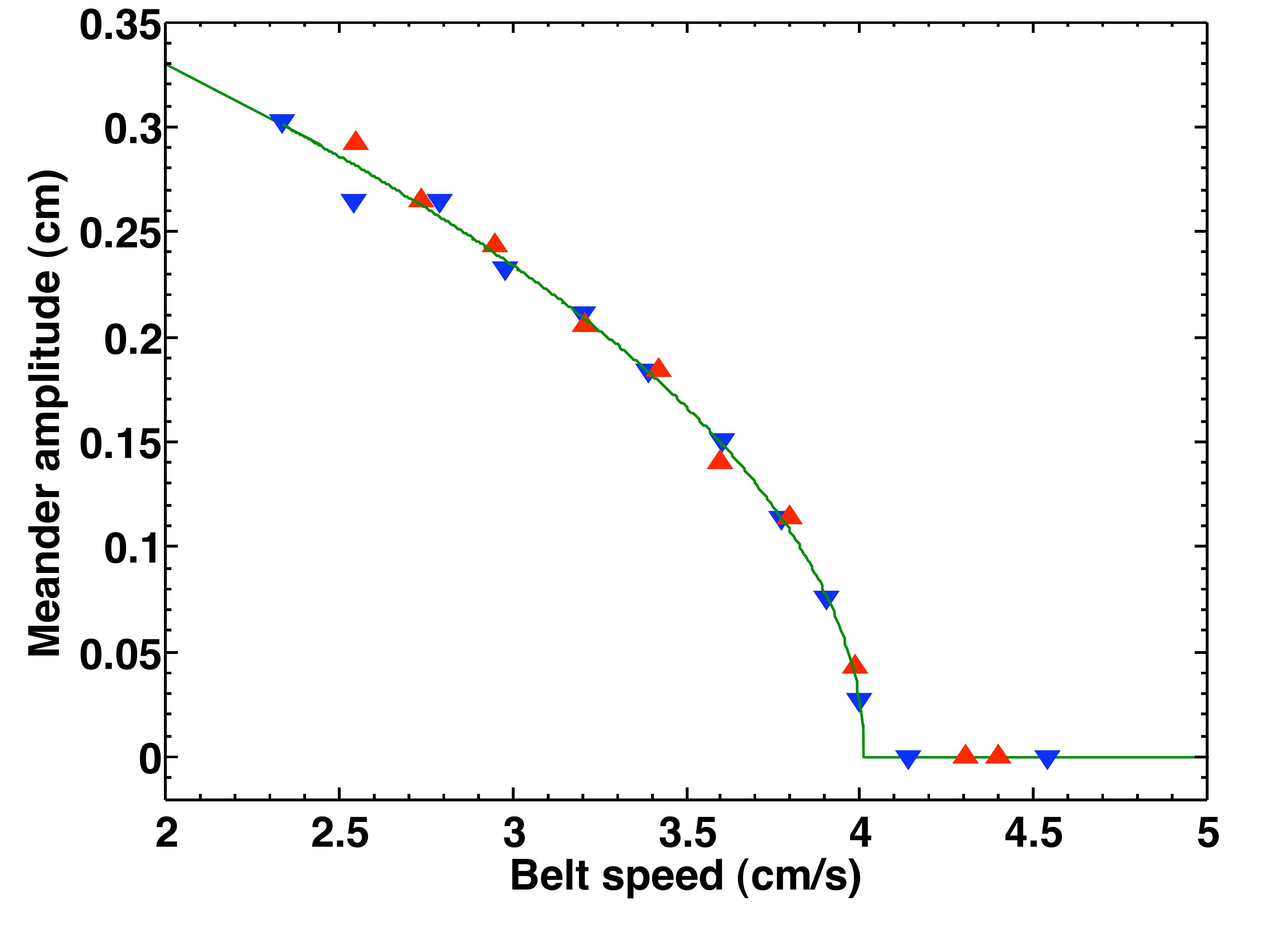}
\caption{(Color online) The amplitude $|A|$ of the meandering motion as a function of the belt speed $U$, for $H = 5.3~{\rm cm}$. The upward (downward) pointing triangles indicate data taken with increasing (decreasing) $U$.  The solid line is a fit to Eqn.~\ref{fit} with $U_c = 4.01~{\rm cm/s}$ and $\mu = 4.62~{\rm cm}^{-2}$.  No higher order terms were required. The smallest nonzero amplitudes here are about half the width of the thread.
%  [[Add comment about thread thickness near onset for comparison with these amplitudes. The text says something about what could be measured but it would be helpful to repeat here that the typical thread thickness is 0.03 cm if that is true. JRL]]
%
%  Done.  SM
%
%
}
\label{amp_vs_speed}
\end{figure}

 We fit the amplitude data to a phenomenological amplitude model of the Landau form
\begin{equation}
%
% Jon wants \lambda here: I think we should stick with epsilon, which
% is deeply entrenched in the physics consciousness. OK - JHPD.
%
\tau {\dot A} = \epsilon A - \mu A |A|^2 + {\rm h.o.t.},
%
%   g changed to mu
%
\label{fit}
\end{equation}
with $\epsilon = (U_c - U)/U_c$. $\tau$ is a time scale related to the linear growth rate.  We consider the steady state of this equation, so $\tau {\dot A}=0$ and use the fit 
% check definition of epsilon === this is right
to determine $U_c$ and the saturation coefficient $\mu$.  
%
%  g is also the standard letter for this coefficient, and I don't think anybody is going to think its
%  the gravitational acceleration.  SM
%
%   It got changed to mu anyway.
%
%
Fig.~\ref{amp_vs_speed} shows such a fit.  The fits were best near the midrange of the heights $H$, declining in quality for small $H$ due to sparsity of data and for larger $H$ due to the growth of the higher order terms.  Near $H \approx 5~{\rm cm}$, the amplitude has an essentially perfect square-root dependence $|A|=\sqrt{\epsilon/\mu}$, as shown in Fig.~\ref{amp_vs_speed}.  For $H > 7.0~{\rm cm}$, fitting was no longer possible, and we simply estimated $U_c$  by bracketting the onset of multifrequency meandering.

\begin{figure}
\includegraphics[width=8cm]{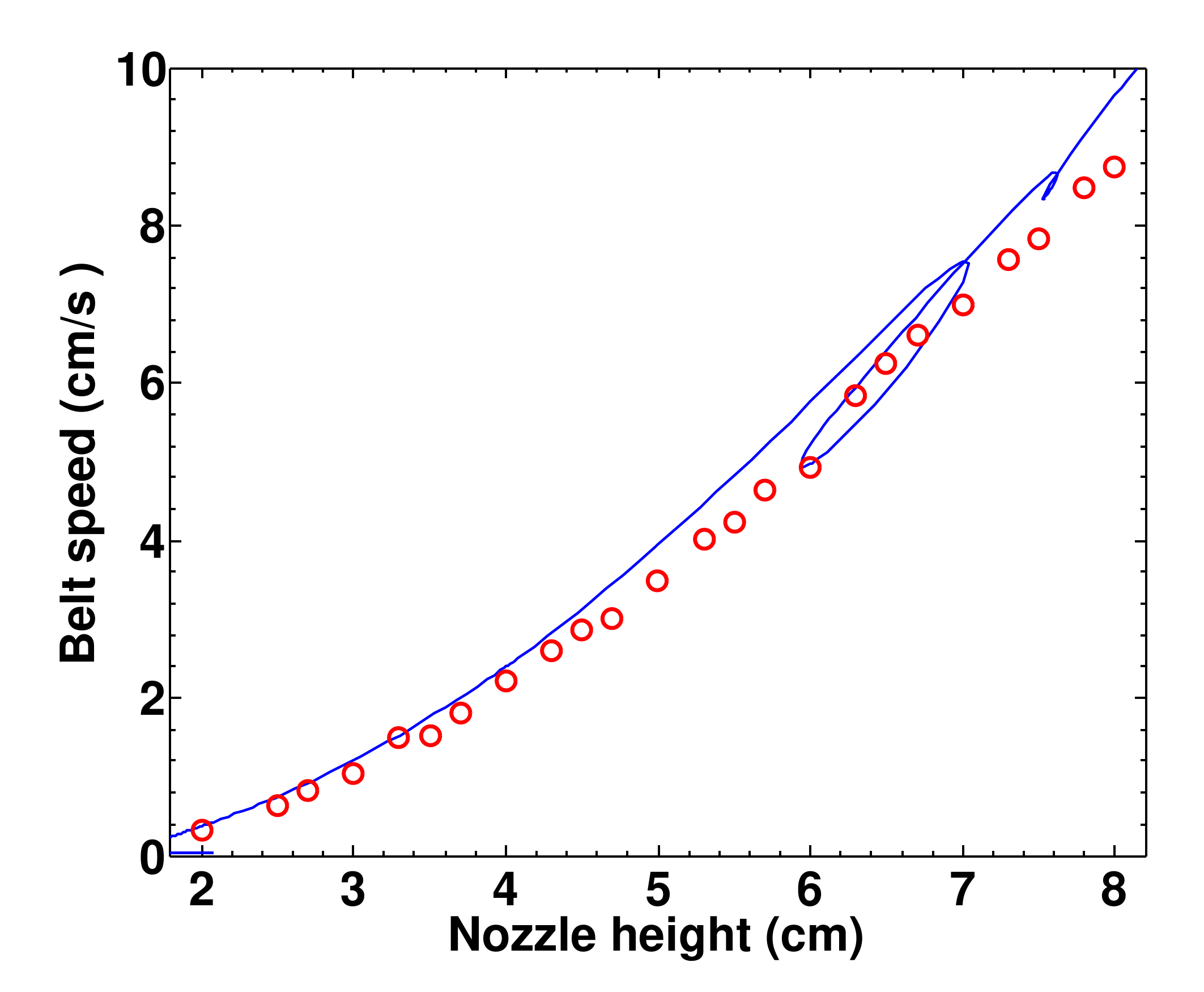}
\caption{(Color online) The critical belt speed $U_c$ vs $H$, extracted from fits to amplitudes like the one shown in the previous figure.  The solid line is calculated from the linear stability analysis of Ref.~\cite{meander_linear_stability}, with no adjustable parameters.}
\label{Uc_Hc}
\end{figure}

Fig.~\ref{Uc_Hc} shows the critical belt speed $U_c$ as a function of the nozzle height $H$, compared to the prediction of the linear stability theory~\cite{meander_linear_stability}. There are no adjustable parameters in the theory. The backward looping portions of the theoretical curve mark the inertial-gravitational regime where multifrequency coiling would exist in the pure rope coiling case.  Since our experimental protocol was to fix $H$ and vary $U$, we would expect to observe the upper envelope of this curve.  The agreement is very satisfactory.

We have no immediate explanation for the slight systematic discrepancy by which the theory overestimates $U_c$ by an amount that increases with $H$.  Presumably it is due to some unaccounted for physical effect, such as air drag or surface tension effects at the contact point between the thread and the belt.  For large $H$, the thread is very sensitive to air currents and the rapidly moving belt inevitably disturbs the air nearby.
%entrains some air.

\begin{figure}
\includegraphics[width=8cm]{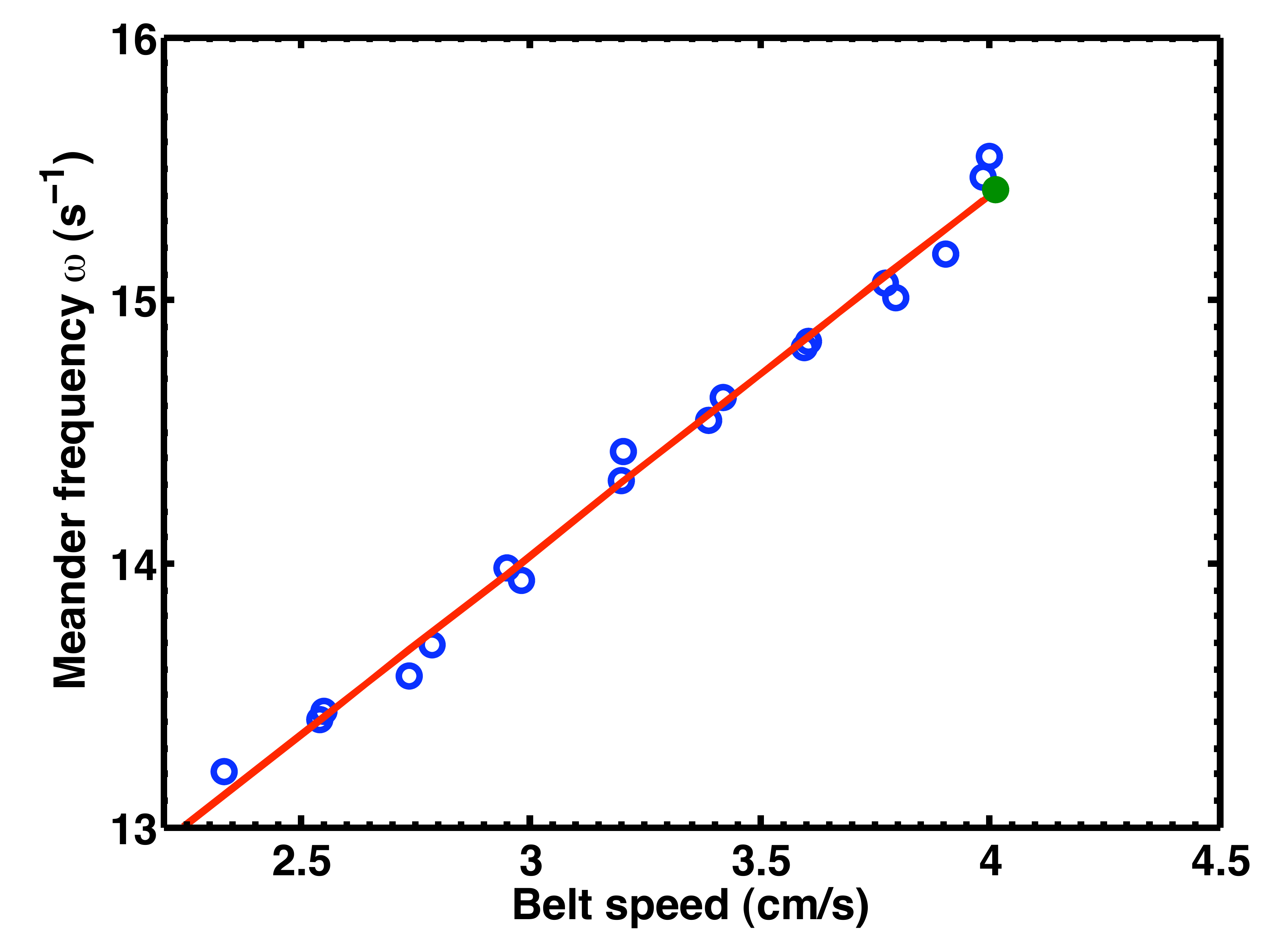}
\caption{(Color online) The angular frequency $\omega$ of the meandering motion (open circles), as a function of the belt speed $U$, for $H = 5.3~{\rm cm}$.  The solid line is a fit to a linear variation of $\omega$ {\it vs.} $U$, 
%with slope $12.5$~cm,   divide by 2 pi  just take this out
consistent with a Hopf bifurcation. The solid circle marks the extrapolation to the critical value $\omega_c$.  }
% If this really is omega then the units are s^{-1}. Same comment in fig 7. JRL
%
%   fixed  SM
%
\label{omega_vs_speed}
\end{figure}

\begin{figure}
\includegraphics[width=8cm]{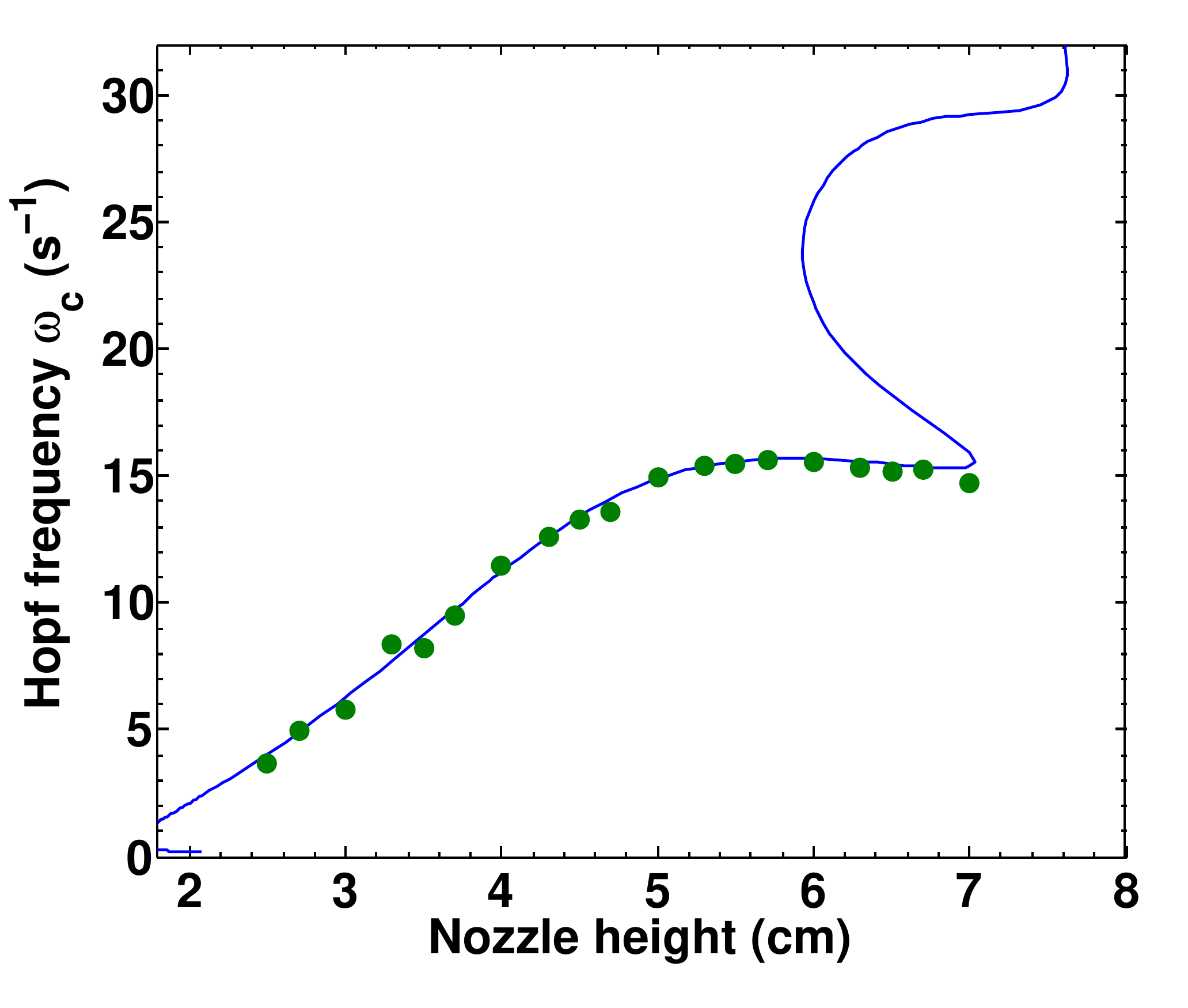}
\caption{(Color online) Measurements of the Hopf frequency $\omega_c$ (solid circles)  as a function of the nozzle height $H$.  The solid line is the theoretical result from the linear stability calculation of Ref.~\cite{meander_linear_stability}.  The reversals in the theoretical curve mark the multifrequency regime. }
\label{omegac_vs_H}
\end{figure}

The meander frequency $\omega$ can also be easily obtained from the measurements of the wavelength $\lambda$ of the meander pattern on the belt.  The angular frequency is simply $\omega = 2 \pi U / \lambda$.  We find that $\omega$ decreases linearly with $U$,  so that it is straightforward to extrapolate to obtain $\omega_c$ as shown in Fig.~\ref{omega_vs_speed}.  The functional forms of the amplitude and frequency of meandering shown in Figs.~\ref{amp_vs_speed} and \ref{omega_vs_speed} demonstrate  that the onset of single frequency meandering is very well described as a forward Hopf bifurcation with Hopf frequency $\omega_c$.  This onset frequency is also predicted by linear stability theory~\cite{meander_linear_stability}; Fig.~\ref{omegac_vs_H} shows the Hopf frequency as a function of $H$.  Here the agreement with the theory is excellent, and again there are no adjustable parameters.

It is interesting that the frequency measurements track the lower branch of the multifrequency regime, right to the end of its existence.  This is a consequence of the experimental protocol in which $H$ is fixed.  Presumably the other frequency branches could be accessed by varying both $H$ and $U$, or perhaps by making finite perturbations to the state of the thread.  The  frequency branches have been observed in pure rope coiling experiments~\cite{maleki_PRL, mult_states}.  The lower branch and the observation of single frequency meandering both end near $H=7.0~{\rm cm}$.  Above this value, we find mostly
% only
 multifrequency meandering at onset, as shown in Fig.~\ref{states}.

%  bigger fonts post-referee

%
% Lister:  July 20 e-mailed version of this section 
%

\subsection{Meandering amplitude saturation:\\ a simple model}
\label{amp_model}

\begin{figure}
\includegraphics[width=8cm]{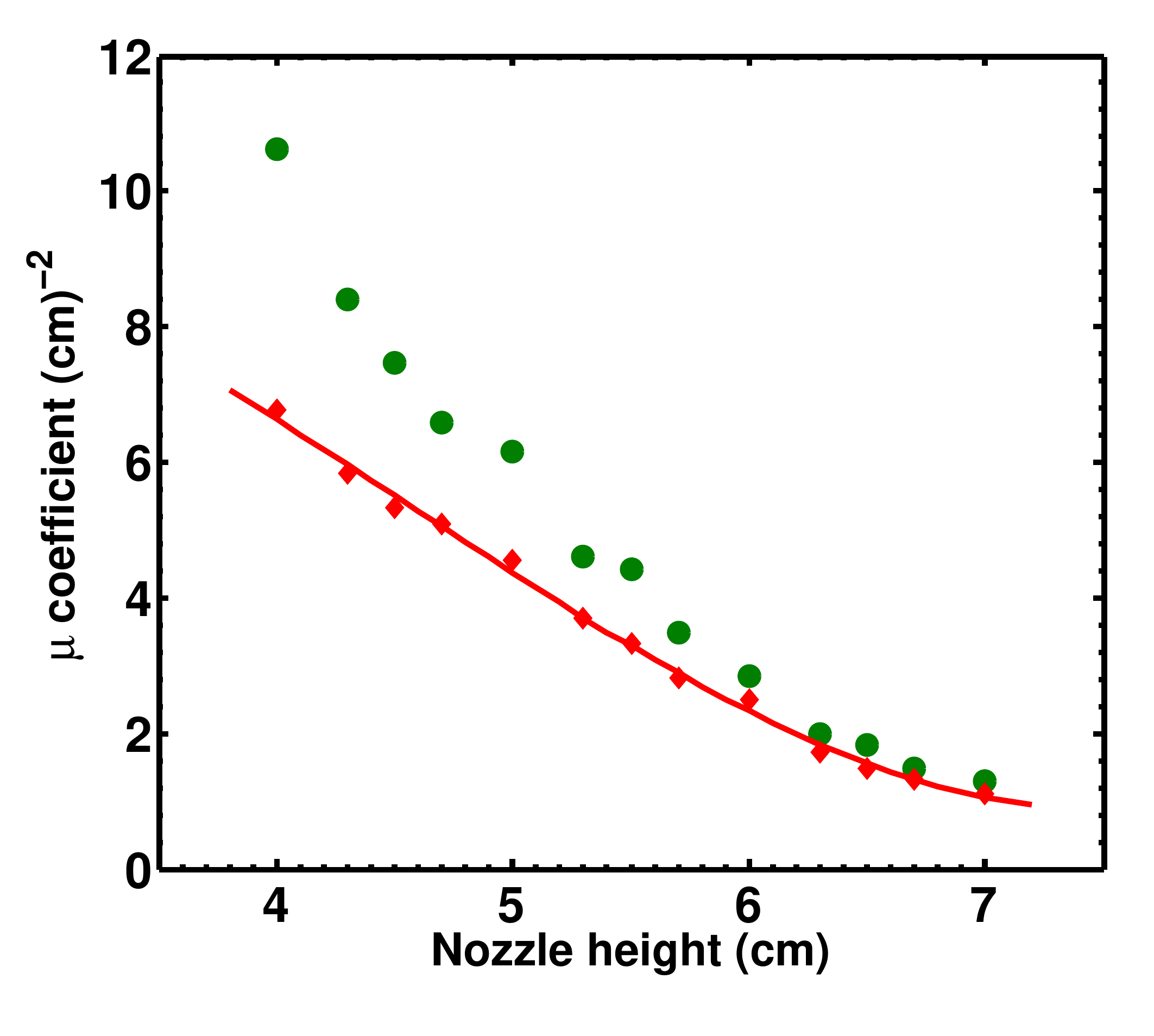}
\caption{(Color online) The nonlinear saturation coefficient $\mu$, as a function of the nozzle height $H$. The solid circles are the values deduced from fits to Eqn.~\ref{fit}.  Diamond symbols show the kinematic estimate $\mu \approx (\omega_c / 2 U_c)^2$ for each height, discussed in the text. The solid line is a cubic polynomial model of the kinematic estimates. }
\label{g_vs_H}
\end{figure}

The fits to the amplitude of meandering also give a quantity that is
not predicted by linear theory, namely the coefficient $\mu$ of the
cubic nonlinear term in Eqn.~\ref{fit}.  Fig.~\ref{g_vs_H} shows the
experimental result.   We find that $\mu$ is a decreasing function of
$H$. The coefficient $\mu$ controls the
dependence of the saturated amplitude on the belt speed.
It can be predicted by the following argument, which is based on
symmetry and the assumption of a constant contact-point speed.

We employ coordinates in the lab frame with $x$ longitudinal to the
belt, $y$ transverse to it and $z$ up. Ribe {\it et
  al.}~\cite{meander_linear_stability} observed that the linear
perturbation equations about a steady stretched catenary decouple into
disjoint systems for perturbations in the plane $(x,z)$ of the
catenary and those out of the $(x,z)$-plane. 
%\marginpar{**} [[This is a consequence of symmetry and, specifically,
%isotypic decomposition, but you might not want to say that...JHPD.
%Agreed about the symmetry, but probably no need to go into it when the perturbation equations are on record. Relocated `further' in following sentence. JRL]]
%OK - JHPD.
Meandering is an out-of-plane perturbation at linear order. We observe that
the reflectional symmetry of the catenary further 
% this alludes to Jon's point 
implies that the nonlinear perturbation equations are invariant to a change in the sign of all the out-of-plane variables. Hence if we perform a series expansion for the weakly nonlinear form of meandering near onset $(A \ll 1)$ then all the in-plane variables are even functions of $A$ and all the out-of-plane variables are odd functions of $A$. In particular, the contact point $(x(t),y(t))$ between thread and belt must have the form
\begin{eqnarray}
%  y(t) &=& A \sin(\omega t) + O(A^3)\\
%  x(t) &=& x_0+B \sin(2 \omega t+\phi)+O(A^4),
%
% SWM:  I made these cosines, in order to maybe use real parts later
%
% JHPD: little adjustments made here to agree with what now follows later.
%
 y(t) &=& A \sin(\omega t) + O(A^3),          \nn \\
 x(t) &=& x_0+B \cos(2 \omega t+\phi)+O(A^4), \label{xy_AB}
  \end{eqnarray}
where $B=O(A^2)$, $\omega=\omega_c+O(A^2)$, $\phi$ is a constant
phase, and $x_0$ is the unperturbed displacement plus an $O(A^2)$
correction.  Thus, the absolute speed $V(t)$ of the contact point 
%
%relative to the belt 
% - I think V(t)=ds/dt is absolute speed, not relative to the belt. JHPD
%
is given by
%
% SWM:  In order to use real parts later there are some little sign changes here
%
%  V^2 &=& (\dot x-U)^2+\dot y^2 \nonumber \\
%    &=& U^2 - 4B \omega U \cos(2 \omega t+\phi) \nonumber \\
%      && ~~ + A^2 \omega^2 [ 1+\cos(2 \omega t) ]/2 +O(A^4)
\begin{eqnarray}
        V^2 &=& (\dot x-U)^2+\dot y^2, \nonumber \\
      &=& U^2 + 4B \omega U \sin(2 \omega t+\phi) \nonumber \\
       && ~~ + A^2 \omega^2 [ 1+ \cos(2 \omega t) ]/2 +O(A^4).
   \label{speedV}
   \end{eqnarray}

We now make the kinematic assumption that $V$ is constant and always equal to the critical belt speed $U_c$. This can be motivated either by the empirical observation in Ref.~\cite{meander_linear_stability} that the steady speed of coiling is remarkably close to the belt speed at marginal stability, or by an asymptotic argument that the amount of stretching in the bending boundary layer of a slender thread is negligible~\cite{blount_in_progress}. Setting $V = U_c$ in Eqn.~\ref{speedV} gives
\begin{equation}
  U_c^2 = U^2 + A^2 \omega^2/2  +O(A^4),
  \label{A_pred}
     \end{equation}
\begin{equation}
  B =  A^2 \omega/(8U) +O(A^4),~~~{\rm and}~~~\phi=\frac{\pi}{2}+O(A^2).
  \label{Bphi_pred}
     \end{equation}
%\marginpar{**}
%[[ could have $\pm A^2 \omega/(8U)$ and then $\phi=\pm
%\frac{\pi}{2}+O(A^2)$ here. JHPD. In (6b) conventional to take $B>0$
%and let $\phi$ cover all possibilities. Hence no need for
%$\pm$. JRL]] OK - JHPD
We rearrange Eqn.~\ref{A_pred},
\begin{equation}
  A^2 = 2(U_c^2-U^2)/\omega^2 \approx (4U_c^2/\omega_c^2)[(U_c-U)/U_c],
  \end{equation}
and compare with Eqn.~\ref{fit} to predict
\begin{equation}
  \mu=(\omega_c/2U_c)^2.
  \label{g_pred}
  \end{equation}

It is evident from Fig.~\ref{g_vs_H} that this purely kinematic prediction of $\mu$  works remarkably well, particularly at larger heights.  In effect, the buckling part of the thread behaves like an inextensible rope being played out onto the moving belt, though there is still extension in the falling part closer to the nozzle. The greater discrepancy between observation and kinematic prediction at lower heights is likely due to weaker scale separation between the falling and buckling parts of the thread.

%
%  Add a brief comment on real ropes here SM
%
It would be interesting to examine the motion of an actual inextensible, flexible rope falling on the moving belt.  Recently, the classic liquid rope-coiling instability has been compared to a real rope in just this way~\cite{real_ropes}.  Presumably, the kinematic condition on the rope at the point of contact with the surface of the belt would have to be enforced by their frictional interaction.

The above kinematic assumption also makes predictions for the amplitude $B$ and
phase $\phi$ of a small, frequency $2\omega$, longitudinal
oscillation. The existence of this oscillation can be discerned in the
spacetime plot of meandering viewed from the side, shown in Fig.~\ref{side_view}(a), which is
discussed in more detail below. The experimental resolution of $B$ is however insufficient to make a %systematic 
quantitative comparison to theory.

The form of Eqn.~\ref{xy_AB} is justified from consideration of the
full amplitude equations, which are presented in Section~\ref{sec:ampeqns}.

\subsection{Beyond meandering: the figure eight and translated coiling states}
\label{beyond}

For $H \le 5.5$~cm, the meandering state gives way at a non-hysteretic bifurcation to the figure eight state shown in Fig.~\ref{state_photos}c.  The figure eight state is, in its turn, unstable to the translated coiling state at lower $H$.  This state links directly to pure fluid rope coiling at $H=0$.  It is difficult to extract quantitative information about these states from photos of the ``stitch pattern'' on the belt.  However, by viewing the thread from the side, the $x(t)$ and $y(t)$ components of motion can be measured.  We accomplished this by positioning a small mirror at $45^\circ$ to the belt, so that a camera could image both the longitudinal and the transverse components at a point close to the location of maximum curvature of the thread, which was a few mm above the surface of the belt. 
%[[[could have a picture for this, I suppose]]] 
%
% We have a lot of figures already.
%
 Taking a single horizontal video line from such an image and plotting it in time gives spacetime images like those in Fig.~\ref{side_view}.

In these images, it is apparent that the meandering state consists of essentially a single frequency $\omega$ in $y$, with a small component at  $2 \omega$ in $x$ with amplitude $B \ll A$, as assumed in Eqn.~\ref{xy_AB} above.  The bifurcation to the figure eight state clearly involves the excitation of a large $2 \omega$ mode in $y$, which mixes with the $\omega$ mode in  $x$ to produce the figure eight pattern.  At the bifurcation to translated coiling at lower $H$, the $2 \omega$ mode amplitude becomes small again, and the state is due to $x$ and $y$ oscillations at $\omega$ which are nearly $90^\circ$ out of phase.

These qualitative observations suggest that the meandering - figure eight - translated coiling series of transitions, and by extension all of the states of motion of the thread, could be understood in the framework of complex amplitude equations for modes with a few frequencies and their harmonics.  These frequencies are probably closely related to the frequencies of pure coiling, including those in the multifrequency regime.

%
%  Long insert from Jon's version:
%
%\subsection{Amplitude equations}   I think this should be all part of
%the same section - OK, JHPD.
% 
% This section has become longer in later versions. Deserves its own
% section now? JHPD.

\section{Amplitude equations and symmetry}
\label{sec:ampeqns}

\begin{figure}
\includegraphics[width=8cm]{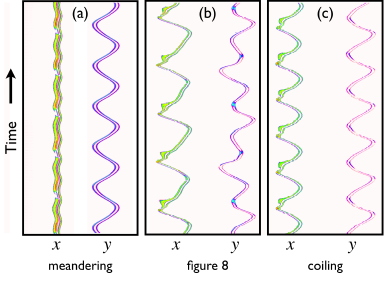}
\caption{(Color online) Spacetime views of the side of the thread, with time running upward.  $x$ is a coordinate longitudinal to the belt, while $y$ is transverse.  (a) shows the meandering state, where almost all the motion is transverse.  (b) shows the figure eight state in which two harmonics of the basic frequency are clearly visible in the longitudinal motion. (c) shows translated coiling, in which one main harmonic appears, with a phase difference between $x$ and $y$.}
\label{side_view}
\end{figure}

We now discuss a more abstract approach to understanding the ``stitch
patterns'' through
the machinery of equivariant bifurcation theory~\cite{GSS88}.
This theory offers a formal way to derive
appropriate amplitude equations,
starting from symmetry considerations alone. Such a qualitative
approach offers a complementary perspective on the dynamics to that
gained from fluid mechanical treatments. It also provides
insight into the overall organisation of the bifurcations between
these highly symmetric states.
% Jon took the following editorial comment out, but I like it ... OK, JHPD.
Given the rich phenomenology of this system, it seems wise
to follow this route rather than attempt to account for all the states
{\it de novo} from fluid mechanics alone. Although a rather large number of amplitudes may be
required, this approach is still much simpler than the alternative.  In principle, the full fluid mechanical description could be systematically projected onto a few modes, and the coefficients of the amplitude equations calculated.  Alternatively, the coefficients could be deduced directly from experimental data of the kind shown in Fig.~\ref{side_view}.

We conjecture that the dynamics of the thread is close to the
axisymmetric case of pure rope coiling, so that the motion of the belt
can be treated as a perturbation. The moving belt breaks the
rotational symmetry while preserving a reflection symmetry in a
vertical plane containing the direction of travel of the belt. In
bifurcation-theoretic terms, the onset of meandering from the steady
catenary state is a Hopf bifurcation with weakly-broken $O(2)$
symmetry \cite{DK91}. We denote the frequency at the Hopf bifurcation
by $\omega$. To describe the secondary bifurcation in which
the figure eight states appear, a minimal description must include
amplitude equations for a mode with frequency $2\omega$. 
We argue in the Appendix that the
%these
 physical symmetries provide enough
constraints to determine the form of the relevant amplitude
equations. In section~\ref{sec:extrasymmetry} we discuss the effect of
a further symmetry property which, while certainly not an
%
%  add word exact post-referee
%
 exact symmetry of the
physical problem, appears to 
%be a property of our
% abstracted bifurcation problem and which may 
help explain certain intriguing aspects of the dynamics of the viscous thread.

We propose an abstracted description of the fluid mechanical problem 
in which we focus on the position of the contact point of the viscous
thread on the belt. We express the contact position of the viscous
thread in the lab frame $(x(t),y(t))$ as the complex position
coordinate
\begin{eqnarray}
p(t) &=& x(t) + i y(t) \nn \\
  &=& A_+(t) \e^{i \omega t} + A_-(t) \e^{-i \omega t}\nn \\
  & & + B_+(t) \e^{2i \omega t} + B_-(t) \e^{-2i \omega t},
  \label{4amps}
% Note: & & + makes much better sense than & + & in the line above. JHPD.
%
%  SWM  experimenting with new notation. Changed to new notation, JHPD.
%
%
 %    &=& A_+(t) \e^{i \omega t} + A_-(t) \e^{-i \omega t}\nn \\
%  &+& B_+(t) \e^{2i \omega t} + B_-(t) \e^{-2i \omega t}, 
\end{eqnarray}
%
%  With this notation, As are always at omega, and Bs are always at 2 omega
% subscript + means exp(+i omega t) and - means exp(- i omega t) etc.
%
%
%  Its still not perfectly clear how (complex) A_1 and A_- are related
%  to (real) A we had before. its not simple. 
%
%  Should be now - see new text added after display of amp eqns and
%  discussion of eps_1, eps_2 and eps_b - JHPD.
%
%
%
where $A_+$, $A_-$, $B_+$ and $B_-$ are the complex amplitudes of four
rotating wave modes.
By symmetry, the generic amplitude equations describing
the dynamics, truncated at cubic order, take the form
\begin{eqnarray}
\dot{A}_+ & = & A_+ \left(\epsilon_1 + i \omega_1
%\dot{A}_+ & = & \epsilon_1 A_+ 
%+ \sum_{j=\pm} \left( a_j |A_j|^2 + b_j |B_j|^2\right)A_+ \nn \\ 
% Just to point out that the above sums are a bit clunky: 
% we've now implicitly defined coeffs a_+ and a_- etc. JHPD.
%
% Actually the sum above was going to get hugely complicated to write
% compactly and CORRECTLY for the other equations. Revert to
% (slightly) longer but accurate notation.
+ a_1 |A_+|^2 + a_2 |A_-|^2 \right. \nn \\
& & \left. + b_1 |B_+|^2 + b_2 |B_-|^2 \right) + s_1 \bar{A}_- B_+ B_- \nn \\
& & + \epsilon_b ( e_1  \bar{A}_- + p_1 \bar{A}_+ B_+ \nn \\
& & + q_1 \bar{A}_+ \bar{B}_- + p_3 A_- B_+ + q_3 A_- \bar{B}_-), \label{eqn:amp1} 
  \end{eqnarray}
  %\\
%& & \nonumber \\
%
%
%
% %%  2 "minus subscripted"  equations removed here post-referee
%
%
%
%\begin{eqnarray}
%\dot{A}_- & = & A_- \left(\epsilon_1 - i \omega_1
%+ \bar{a}_1 |A_-|^2 + \bar{a}_2 |A_+|^2 \right. \nn \\
%& & \left. + \bar{b}_1 |B_-|^2 + \bar{b}_2 |B_+|^2 \right) + \bar{s}_1
%\bar{A}_+ B_+ B_- \nn \\
%& & +\epsilon_b ( \bar{e}_1 \bar{A}_+ + \bar{p}_1 \bar{A}_- B_-  \nn \\
%& & + \bar{q}_1 \bar{A}_- \bar{B}_+ + \bar{p}_3 B_- A_+ + \bar{q}_3 A_+ \bar{B}_+), \label{eqn:amp2}
%    \end{eqnarray}
%
%
%
    % \\
%& & \nonumber \\
%
\begin{eqnarray}
\dot{B}_+ & = & B_+ \left(\epsilon_2 + i \omega_2
%\dot{B}_+ & = & \epsilon_2 B_+ 
%+ \sum_{j=\pm} \left(c_j |A_j|^2 + d_j |B_j|^2 \right)B_+ \nn \\
% Again, too complicated. JHPD.
+ c_1 |B_+|^2 + c_2 |B_-|^2 \right. \nn \\
& & \left. + d_1 |A_+|^2 + d_2 |A_-|^2 \right) + s_2 \bar{B}_- A_+ A_- \nn \\
& & + \epsilon_b ( e_2  \bar{B}_- + p_2 \bar{A}_-^2 + q_2 A_+^2 + p_4 A_+ \bar{A}_-),
\label{eqn:amp3} 
  \end{eqnarray}
  %\\
%& & \nonumber \\
%
%
%
%  minus subscript equation removed post-referee
%
%
%\begin{eqnarray}
%\dot{B}_- & = & B_- \left(\epsilon_2 - i \omega_2
%+ \bar{c}_1 |B_-|^2 + \bar{c}_2 |B_+|^2 \right. \nn \\
%& & \left. + \bar{d}_1 |A_-|^2 + \bar{d}_2 |A_+|^2 \right) + \bar{s}_2 \bar{B}_+ A_+ A_- \nonumber \\
%& & +\epsilon_b ( \bar{e}_2 \bar{B}_+ + \bar{p}_2 \bar{A}_+^2 +
%  \bar{q}_2 A_-^2 + \bar{p}_4 \bar{A}_+ A_-), \label{eqn:amp4}
%\end{eqnarray}
%\marginpar{JRL}[[Can we take the factor $A_+$ out of the parentheses in (13), etc.?]]
%Done - and rewritten to correct the terms in the \sum bits. JHPD.
and two similar equations for $A_-$ and $B_-$, given in the Appendix.  The coefficients of the nonlinear terms 
in Eqns.~\ref{eqn:amp1} and \ref{eqn:amp3} are in
general complex-valued, and overbars denote complex conjugates.
%As argued in the Appendix, 
%due to the symmetry $h \circ \kappa$ the coefficients are expected to have
%only small imaginary parts. Similarly we expect $\epsilon_1$, $\epsilon_2$
%to have only small imaginary parts, i.e. that the frequency of the
%meandering oscillation does not vary substantially with increasing
%meandering amplitude. In what follows, for clarity, we take
The bifurcation parameters $\epsilon_1$ and $\epsilon_2$
are complicated unknown functions of the dimensionless
groups identified above, or equivalently of the natural experimental
control parameters $U$ and $H$. $\epsilon_1(U_c,H)=0$ at the onset of
meandering, as described above in section~\ref{sec:meanderingonset}.
% and in figure~\ref{amp_vs_speed}.
$\epsilon_2(U_c,H)$ remains negative and describes the linear
damping of the $2\omega$ mode. The parameter $\epsilon_b$, a function
of $U$ with $\epsilon_b(U=0)=0$, controls the strength of the
symmetry-breaking effects of the belt on the thread. 
%The other coefficients may be treated as constant.
%
%  But they are probably not.  We don't really "treat" them at all anyway.
%  Agreed. JHPD.

%The details of the derivation of these amplitude equations can be
%found in the Appendix. 
Eqns.~\ref{eqn:amp1} and \ref{eqn:amp3} 
justify the use of the phenomenological  Eqns.~\ref{fit}
and~\ref{xy_AB}, as we discuss further below.
A full analysis of the bifurcation structure of
Eqns.~\ref{eqn:amp1} and \ref{eqn:amp3} 
%
%  I like these words better:  SM
%
%
is mathematically feasible, but beyond the scope of this paper.
%
%deferred to a future
%paper.  
%
%   I avoid saying anything about papers that may never be written ...  SM
%
In the remainder of this section, we first discuss the meandering
instability, and then the transition from meandering to figure
eights, in the context of these amplitude equations.

The steady catenary solution $A_+=A_-=B_+=B_-=0$ exists for all parameter
values and is stable if $\epsilon_1 \pm \epsilon_b {\rm Re}(e_{1}) <0$.
Experimentally, there is a critical height $H_0 \approx 2.0$~cm
below which no rope coiling instability occurs for $U=0$. It is therefore 
natural to speculate that, for small $U$ and heights close to
$H_0$, we could assume $\epsilon_1 \propto (H-H_0)$ and $\epsilon_b \propto
U$. These scalings in turn
predict a straight-line marginal stability boundary for the onset of
meandering near this endpoint. This regime is difficult to reach
experimentally, however, due to the extremely slow time evolution and
the thickening of the thread relative to its meandering amplitude.

% we didn't really try to do much at this extreme 

%For small $U$ practical difficulties mean that substantial
%discrepancies arise between theory and experiment due to a build-up
%of fluid beneath the slowly coiling thread, changing the effective
%height of the thread. It is also not clear exactly what the relevant
%boundary condition for the numerical solution should be.
% ****  Please could others comment further here if necessary.
% ****

%Meandering states bifurcate from the steady
%catenary and are expected to persist, although not to be stable, 
%down to $U=0$. 

In the axisymmetric case $U=0$, when $\epsilon_b=0$,  there is an
$O(2)$-symmetric Hopf bifurcation at $H=H_0$ in which two oscillatory
branches appear~\cite{GS85}: a rotating wave corresponding to coiling,
and a standing
wave corresponding to meandering. The coiling state
persists under weak symmetry-breaking~\cite{DK91} and evolves
smoothly into the translated coiling state $A_+ \neq 0$,
$A_-=B_+=B_-=0$, observed for small $U$. In
the axisymmetric case, the bifurcating
standing wave has no preferred horizontal
orientation. Under weak symmetry-breaking, two standing waves out of
this family of states survive. One of these states corresponds to
oscillations where the thread lies entirely within the symmetry-plane
of the belt at all times 
%(we refer to this as ``longitudinal meandering'' or ``slumping''), JHPD
% longitudinal oscillations really can't be said to `meander' in a meaningful sense JRL
%(``pure  - all things are pure to the pure in heart. JHPD.
(a ``longitudinal oscillation''), 
%
%\marginpar{**}
%[[Did Chiu-Webster and Lister give this a name? - observed by them at higher
%H,U I believe...JHPD We called them longitudinal oscillations between ourselves, but I don't think we mentioned them in the paper. JRL]]
%
the other (``meandering'') has the combination
spatiotemporal symmetry $\kappa \circ \tau_\pi$
(defined precisely in the Appendix) of reflection across
the belt followed by waiting for a half-period of the oscillation.

The figure eight state also has the symmetry $\kappa \circ \tau_\pi$
and so, to simplify the further discussion of meandering and figure
eights, we restrict
our attention only to solutions that have the $\kappa
\circ \tau_\pi$ symmetry, that is, which have the form
$A_+=-\bar{A}_-\equiv A$ and $B_+=\bar{B}_-\equiv B$. We refer
to the invariant subspace $(A,-\bar{A},B,\bar{B})$ as $\S$;
importantly, $\S$  remains invariant even if higher-order terms are
added to the amplitude equations.

Within $\S$, the amplitude equations simplify 
%(dropping the subscript `$+$')
to become
\ba
%\dot{A}_+ & = & (\epsilon_1 + i \omega_1 -\epsilon_b e_1)A_+ +\epsilon_b
%(p_1 + q_1 - p_3 -q_3) \bar{A}_+ B_+ \nn \\
%& & + (a_1 + b_1) A_+ |A_+|^2 + (a_2+b_2-s_1) A_+ |B_+|^2, \label{eq:fixkt1} \\
%\dot{B}_+ & = & (\epsilon_2 + i \omega_2 + \epsilon_b e_2)B_+ +
%\epsilon_b (p_2 + q_2 - p_4) A_+^2 \nn \\
%& & + (c_1 + d_1 - s_2) B_+ |A_+|^2 + (c_2+d_2) B_+ |B_+|^2. \label{eq:fixkt2}
%
%\dot{A}_+ & = & (\epsilon_1 + i \omega_1 -\epsilon_b e_1)A_+ + i \epsilon_b
%\alpha_1 \bar{A}_+ B_+ \nn \\
%& & + \beta_1 A_+ |A_+|^2 + \beta_2 A_+ |B_+|^2, \label{eq:fixkt1} \\
%\dot{B}_+ & = & (\epsilon_2 + i \omega_2 + \epsilon_b e_2)B_+ +
%i \epsilon_b \alpha_2 A_+^2 \nn \\
%& & + \beta_3 B_+ |B_+|^2 + \beta_4 B_+ |A_+|^2. \label{eq:fixkt2}
%
\dot{A} & = & (\epsilon_1 +i\omega_1 -\epsilon_b e_1)A +i\epsilon_b
\alpha_1 \bar{A}B 
%\nn \\ & & 
+ \beta_1 A|A|^2 + \beta_2 A|B|^2, 
\nn \\ \label{eqn:fixkt1} \\
\dot{B} & = & (\epsilon_2 +i\omega_2 +\epsilon_b e_2)B + i
\epsilon_b \alpha_2 A^2 
%\nn \\ & & 
+ \beta_3 B|B|^2 + \beta_4 B|A|^2. 
\nn \\ \label{eqn:fixkt2}
\ea

The coefficients $e_1,e_2,\alpha_1,\alpha_2,\beta_1,\ldots,\beta_4$
in Eqns.~\ref{eqn:fixkt1} and \ref{eqn:fixkt2} can be
deduced by direct comparison with
Eqns.~\ref{eqn:amp1} and \ref{eqn:amp3}.
Near the initial meandering instability
$\epsilon_1~-~\epsilon_b{\rm Re}(e_{1})=0$, and we may eliminate $B$
adiabatically, or {\it via} a centre manifold reduction, so that to leading
order
$B=-i\epsilon_b \alpha_2 A^2 / (\epsilon_2 + i \omega_2 + \epsilon_b
e_2) + O(A^3)$
%$B=i\epsilon_b \alpha_2 A^2 / (\epsilon_2 + \epsilon_b e_2) + O(A^3)$
and hence
\ba
\dot{A} & = & (\epsilon_1 +i\omega_1 - \epsilon_b e_1)A + \left(
\frac{\epsilon_b^2 \alpha_1 \alpha_2}{\epsilon_2 + i\omega_2 +\epsilon_b e_2}+\beta_1
\right) A|A|^2
%+ O(A^4) % - deleted due to space on the line!
\nn
\ea
(up to $O(A^3)$), which is the usual amplitude equation for a Hopf bifurcation,
justifying the use of Eqn.~\ref{fit}.

Writing $A=R_1 \e^{i \theta_1}$ and $B=R_2
\e^{i \theta_2}$ we find that, due to the time-translation symmetry,
Eqns.~\ref{eqn:fixkt1} and \ref{eqn:fixkt2} 
reduce further, to a three-dimensional set of ODEs for $R_1$,
$R_2$ and the relative phase $\chi=\theta_2-2\theta_1$.
%
%
%
%   removed 3 equations post-referee
%
%
%\ba
%\dot{R}_1 & = & (\epsilon_1-\epsilon_b {\rm Re}(e_1)) R_1 - \epsilon_b
%  R_1 R_2 {\rm Im}\left(\alpha_1\e^{i\chi}\right)
%\nn \\ & & + {\rm Re}(\beta_1) R_1^3 + {\rm Re}(\beta_2) R_1 R_2^2,
%\label{eqn:r12} \\
%& & \nn \\
%\dot{R}_2 & = & (\epsilon_2+\epsilon_b {\rm Re}(e_2)) R_2 - \epsilon_b
%R_1^2 {\rm Im}\left(\alpha_2 \e^{-i\chi}\right) 
%\nn \\ & & + {\rm Re}(\beta_3)R_2^3 + {\rm Re}(\beta_4) R_2 R_1^2, \label{eqn:r22} \\
%& & \nn \\
%\dot{\chi} & = & \omega_2-2\omega_1 + \epsilon_b{\rm Im}(e_2+2e_1) \nn
%\\
%& & + R_1^2 {\rm Im}(\beta_4-2\beta_1) + R_2^2 {\rm
%  Im}(\beta_3-2\beta_2) \nn \\
%& & + \epsilon_b \left(
%\frac{R_1^2}{R_2}{\rm Re}\left(\alpha_2 \e^{-i\chi}\right) -
%2 R_2 {\rm Re}\left(\alpha_1 \e^{i\chi} \right)\right). \label{eqn:chi2}
%\ea
%From Eqn.~\ref{eqn:chi2}, we note that t
There are regions of parameter
space in which time-periodic solutions with a constant
relative phase $\chi$ do not exist. In other regions of parameter
space, there will typically be two possible
constant values $\chi=\chi_\pm$. 
%If the coefficients in Eqns.~\ref{eqn:fixkt1}~-~\ref{eqn:fixkt2} have
%(small) nonzero imaginary parts then the two time-periodic states
%persist but now only have $\chi\approx \pm \pi/2$. 
In general, within $\mathcal{S}$ we recover the form
\ba
y(t) & = & 2R_1 \sin \omega t,  \\
x(t) & = & 2R_2 \cos (2\omega t + \chi), 
\ea
for the location of the tip of the thread, which confirms
Eqn. \ref{xy_AB}.

%

%
%
%  The following paragraph is problematic.  It seems that we do not
%  have a good theory for the emergence of the figure eight state from
%  meandering!!
%
% The following paragraph has been rewritten from scratch - a
% combination of my earlier confusion between bunched-up meandering
% and fig-8 plus the 'extra symmetry' realisation have made all this
% much clearer now.

%LEFT ONLY FOR REFERENCE! JHPD.
%Perhaps surprisingly, theory and experiment both indicate that the
%figure eight state emerges from meandering. In fact, as
%figure~\ref{side_view}(a) shows, the meandering motion always contains a small
%amplitude $2\omega$ component. Theoretically this is due to the terms
%$\epsilon_b (p_2 \bar{A}_-^2 + q_2 A_+^2 + p_4 A_+ \bar{A}_-)$
%in~(\ref{eqn:amp3}) (and similar terms in~(\ref{eqn:amp4})) which show that
%the two primary amplitudes drive a small response in the $2\omega$
%modes, and this occurs \textit{only\/} in the presence of the symmetry-breaking
%induced by the belt motion. As $U$ decreases the amplitude of the
%$2\omega$ modes should grow continuously, but this fails to explain
%the abrupt, but non-hysteretic, appearance of the figure eight shape
%at lower $U$.

If $\epsilon_1>0$ and $\epsilon_2<0$ (as expected for the
onset of meandering when $H<6.0$~cm), and if $\epsilon_b=0$,
we would expect
stable meandering to exist for $0<\epsilon_1<{\rm Re}(\beta_1)
\epsilon_2/{\rm Re}(\beta_4)$, choosing ${\rm Re}(\beta_1)<0$
to ensure that the onset of meandering is supercritical, and ${\rm
Re}(\beta_4)>0$ to ensure the existence of an instability
of meandering to the $2\omega$ mode that produces the figure eight
pattern. At $\epsilon_1={\rm Re}(\beta_1) \epsilon_2/{\rm
Re}(\beta_4)$ there is a
pitchfork bifurcation at which meandering loses stability to
``mixed-modes'' involving the $2\omega$ mode.
For $\epsilon_b \neq 0$ the pitchfork bifurcation becomes imperfect,
with a continuous, but rapid and 
%seemingly 
non-hysteretic,
increase in the amplitude of the $2\omega$
mode in the vicinity of $\epsilon_1={\rm Re}(\beta_1)
\epsilon_2/{\rm Re}(\beta_4)$.
%Fig.~\ref{fig:bifns} illustrates the onset of
%meandering with increasing $\epsilon_1$ (e.g. increasing $H$ at fixed
%$U=1.5$cm/s) followed by the rapid, and non-hysteretic,
%evolution of the meandering into the figure eight pattern.
%\begin{figure}
%\includegraphics[width=8.5cm]{composite1.eps}
%\caption{Illustrative bifurcation diagram for Eqns.~
%  \ref{eqn:r1}~-~\ref{eqn:chi} showing the onset of
%  meandering at $\epsilon_1=0$ and the rapid transition from
%  meandering to figure eights at $\epsilon\approx\beta_1
%  \epsilon_2/\beta_4=0.8$. (a) $\epsilon_1=0.4$;
%(b) $\epsilon_1=0.8$; (c) $\epsilon_1=0.9$; (d) $\epsilon_1=1.0$; 
%(e) $\epsilon_1=1.1$. Coefficients: $\epsilon_2=-1,
%\epsilon_b=0.01,e_1=e_2=1,\alpha_1=0,\alpha_2=-1,
%\beta_1=-2,\beta_2=0.5,\beta_3=-1,\beta_4=2.5$.}
%\label{fig:bifns}
%\end{figure}

Thus,  the sequence of transitions observed
experimentally has a natural interpretation in terms of the mode
interaction described above. 
%
%   This discussion of bunched up meandering is problematic: I think
%  it might be a transient or related to self-sticking.
%
%  It happens at higher H that we said we would not discuss!
%
%  Anyway, we have not discussed it in this paper
%
%Interestingly, this scenario predicts that
%just before the appearance of the figure eight state, the meandering
%should become more pointed, rather than more rounded as is the case in
%the ``bunched-up meandering'' state (which was observed in Ref.~\cite{chu_webster}, but not clearly in the present experiment). 
%
%  We have said nothing about bunched up meanders so far in this paper.
%
%  I will have to check, but I don't think I saw much bunched up meandering   SM
%
%  May need to tweak the wording here
%
%It is not clear that this is
%supported by current experimental results, or by the 
%``constant contact
% point speed'' 
%kinematic assumption of section~\ref{amp_model} that supports the
% formation of bunched-up meanders.
%
%  
%
%

Moreover, the experimental evidence 
indicates a  discontinuous but {\it  not} hysteretic
transition from meandering to
figure eight; from the theory one would expect either that the
transition is indeed abrupt (for example a subcritical bifurcation)
and accompanied by measurable hysteresis, or that the
transition is non-hysteretic but continuous. The resolution of the
exact nature of the transition from meandering to the figure eight state
%
%  words  SM
%
%is one of many features of this problem that we will re-examine
% with a yet more carefully controlled experimental setup. 
will require better controlled experiments to sort out.
Direct measurement of the amplitudes and phases of the $\omega$ and $2\omega$ components
of the pattern should enable this intriguing discrepancy to be
resolved.

%Two features of the bifurcation analysis explain, qualitatively,
%experimental features observed here and by Chiu-Webster \& Lister
%\cite{chu_webster}.

%Firstly, the amplitudes of the mixed-modes are
%determined from roots of a cubic polynomial; set $\chi=\pm \pi/2$ and
%solve Eqns.~\ref{eqn:r1}~-~\ref{eqn:r2} for $R_1$ and $R_2$. Thus,
%for typical nonlinear coefficients, 
%at finite amplitude beyond the meandering instability
%there are curves in the $(\epsilon_1,\epsilon_2)$ (equivalently, the
%$(H,U)$ plane) plane along which a pair of real roots appears and there are a
%saddle-node bifurcations of mixed-mode (either meandering or
%figure-eight) states.

\subsection{An additional model symmetry}
\label{sec:extrasymmetry}

%%% some words added here post-referee:

%A further point of interest in
%the above discussion is that 
The amplitude equations, discussed above, do
not precisely determine the relative phase $\chi$. However,
experimental results (Fig. \ref{state_photos} as well as those of Ref.~\cite{chu_webster}) 
indicate that the values $\chi=\pi/2$ ( the ``bunched-up
meandering'' state discussed in Ref.~\cite{chu_webster}) and near $\chi=-\pi/2$ (figure eight) are consistently
selected. It is
natural to ask whether there is a reason for this, either of an
abstract or a fluid-mechanical nature.

As far as the abstract model problem goes, it turns out that
there is indeed an
additional symmetry that implies $\chi=\pm \pi/2$. This symmetry
(labelled $h$ in the Appendix) is a
$180^\circ$ rotation of the $(x,y)$ plane followed by reversal of the
direction of time. Clearly this is not a symmetry of the original
physical problem: in no sense are we suggesting that falling fluid in
the thread is related to a viscous thread of rising fluid! But in the
bifurcation-theoretic
model,  we ignore the vertical direction, and the time-reversal
applied to the equilibrium state results only in the reversal of the
direction of travel of the belt. Even so, we would expect $h$ to be
only an approximate symmetry since the initial steady catenary state
is not symmetric under a $180^\circ$ rotation. The consequences of the
extra symmetry $h$ would be to force the imaginary parts of all
coefficients in Eqns. \ref{eqn:fixkt1} and \ref{eqn:fixkt2} to be real.
%
%
%  changed post-referee
%In this case, Eqns.~\ref{eqn:r12}~-~\ref{eqn:chi2} become
%
In this case, the ODEs for $R_1$,
$R_2$ and the relative phase $\chi=\theta_2-2\theta_1$ become
\ba
\dot{R}_1 & = & (\epsilon_1-\epsilon_b e_1) R_1 - \epsilon_b \alpha_1
R_1 R_2 \sin \chi + \beta_1 R_1^3 + \beta_2 R_1 R_2^2, \nn \\ \label{eqn:r1} \\
\dot{R}_2 & = & (\epsilon_2+\epsilon_b e_2) R_2 + \epsilon_b \alpha_2
R_1^2 \sin \chi + \beta_3 R_2^3 + \beta_4 R_2 R_1^2, \nn \\ \label{eqn:r2} \\
\dot{\chi} & = & \epsilon_b \left( \frac{\alpha_2 R_1^2}{R_2} -
2 \alpha_1 R_2 \right) \cos \chi. \label{eqn:chi}
\ea
Hence, from Eqn.~\ref{eqn:chi}, time-periodic solutions with a constant
relative phase exist only for $\chi=\pm \pi/2$.
If the coefficients in Eqns.~\ref{eqn:fixkt1} and \ref{eqn:fixkt2} have
(small) nonzero imaginary parts then the two time-periodic states
persist but now only have $\chi\approx \pm \pi/2$. There is some
experimental evidence that the bunched-up meandering state observed in 
Ref.~\cite{chu_webster} has $\chi \approx \pi/2$, without
having exact equality.

%Synthetic reconstructions of the
%motion for different values of $\chi$ show that the bunched-up
%meandering state corresponds to the time-periodic solution with
%$\chi\approx\pi/2$, and the ``figure eight'' state coresponds to $\chi
%\approx-\pi/2$.
%
%\marginpar{**}
%[[ Should we add a synthetic figure here to show this?? JHPD. CWL
%commented on the opposite phase of figure 8. JRL]]. Added references
%to CWL paper. JHPD.

More generally,
% speaking,
Eqns.~\ref{eqn:r1}~-~\ref{eqn:chi} are equivalent
to the amplitude equations of the well-studied $1:2$
steady-state mode interaction \cite{D86,AGH87,PJ88,DPP04}.
The meandering and figure eight states correspond to
``mixed-mode'' equilibria in 
%these papers 
this language, since both states
contain non-zero
amounts of both the $\omega$ and $2\omega$ modes. However, the analysis in
Refs.~\cite{D86,AGH87,PJ88,DPP04} concentrates on cases where the quadratic
terms have order unity coefficients and strongly influence the
dynamics near $R_1=R_2=0$. In contrast, in the present case, we are
interested in the regime $\epsilon_b \ll 1$ where we see
from Eqns.~\ref{eqn:r1}~-~\ref{eqn:chi} that the quadratic terms fix
the relative phase $\chi$ but only weakly affect the amplitudes $R_1$
and $R_2$.

Having derived the form of the amplitude equations on symmetry
grounds, an obvious next step would be to compute or measure the undetermined
coefficients 
%contained 
in them. 
%From this
%viewpoint, the physics of the system is contained entirely in the values of 
%these coefficients. 
It is straightforward to extend Eqns.~\ref{eqn:amp1} and \ref{eqn:amp3}
 to account for higher frequencies, although the
number of terms, even at cubic order, grows quite rapidly.
Projecting the full dynamics of the thread onto these equations serves
to simplify the problem to finding the coefficients
%$a_1, a_2 ... b_1,b_2 ...$, 
% coeffs aren't called exactly this anymore - omit without loss of
% clarify. JHPD.
and how they depend on the nondimensional experimental
parameters $H(g/\nu^2)^{1/3}$,
%  check that this is the right expression for the dimensionless U
%
% SWM:  I guess its not.  John says:
% 
%I didn't see why U would be made dimensionless w.r.t. (g nu^4/H^3)^{1/6}. Viscous thread theory says g H^2/\nu is a more obvious choice.
%
%
%$u = U(H^3/(g \nu^4))^{1/6}$,  
%
$U(g H^2/\nu)$ 
%
% What did we decide about this scaling?
%
and $\Pi_1$, $\Pi_2$ and $\Pi_3$ given by 
%Eqn.~\ref{pi_1_pi_2}
Eqns.~\ref{pi_1}-\ref{pi_3}.
In principle, the entire state diagram shown in Fig.~\ref{states} may
be understood in this way.

%
% already said this enough times
%
%In summary, our expectation is that a low-order model of this
%kind should help provide detailed insights into the nonlinear dynamics while
%avoiding the complexity of solving the full fluid-mechanical
%time-dependent boundary-value problem. 

\section{conclusions}
\label{conc}

%%Make a succinct summary here.
%
%%

%% A first, very brief, stab at conclusions. JHPD
In this paper we have reported a detailed experimental survey of the
``stitch patterns'' observed in a more precise version of the
``fluid-mechanical sewing machine''~\cite{chu_webster}.   We studied, in particular, the onset of the meandering state and measured its critical belt speed, Hopf frequency and amplitude saturation.  Our theoretical understanding and interpretation of this instability of the catenary thread comes from three sources: a detailed numerical
solution of the linear stability problem~\cite{meander_linear_stability}, 
% a subtle
an \textit{ad hoc}
argument which used the assumption that the contact point of the
thread with the belt has a constant velocity, and the construction of the full 
amplitude equations on symmetry grounds.

%** Now say something about each of these in turn.

We found that the onset of meandering is very well described as a Hopf bifurcation with the critical parameters $U_c$ and $\omega_c$ accurately predicted by a linear stability analysis of the fluid equations~\cite{meander_linear_stability}.  Surprisingly, the nonlinear saturation of the meandering motion could be mostly accounted for by simply assuming that the fluid thread behaves as an inextensible rope being played out onto the belt.  This is the case despite the considerable buckling and stretching of the real thread.  Finally, the full amplitude equations, derived from symmetry considerations independent of the fluid mechanical details, provide a systematic way of organizing and interpreting the otherwise rather complex motion of the thread. 

From the full amplitude equations we were able to explain, at least
in outline, the existence of the figure eight
bubble that occurs over a substantial part of the
state diagram for smaller $H$, as seen in Fig.~\ref{states}.  It seems likely that 
this state is a manifestation of a $1:2$
steady-state mode interaction \cite{D86,AGH87,PJ88,DPP04}.  In general, such a mode interaction is well known to generate complex dynamics in certain cases, such as when
% (the real parts of) 
$\alpha_1$ and $\alpha_2$ in Eqns.~\ref{eqn:r1}~-~\ref{eqn:chi} have opposite signs.  In this case, one expects 
quasiperiodic oscillations and robust heteroclinic
cycling. Physically, such dynamical complexity would be most easily
observed in the intermittent evolution of the relative phase variable $\chi$ on a
timescale that is slow compared to the basic meandering oscillation
period.  Future, improved experiments could look for such subtle effects.  

An interesting question that needs to be addressed more fully, both in
experiment and theory, is the variation of the oscillation frequency
as the belt speed is reduced. Ribe {\it et. al.}
\cite{meander_linear_stability} compared the frequency at the onset of
meandering to the frequency of finite amplitude ``pure'' coiling (when $U=0$)
at the same height $H$ and found excellent agreement. Our results, however,
indicate that the meandering frequency decreases substantially from that
at onset as the belt speed decreases, as shown in Fig.~\ref{omega_vs_speed}.  This decrease is exactly what is expected from the linear amplitude dependence of the frequency following a Hopf bifurcation.  But how can this be reconciled with the requirement that the frequency must also agree with the pure coiling frequency again when $U$ is ultimately reduced to zero? Is it the case that the frequency reaches a local minimum as $U$ is decreased, before increasing continuously back towards pure coiling, or are there discontinuities?  What is the effect of the intervening figure eight state on the fundamental frequency?

For larger values of $H$, the amplitude equation model would have to be extended to allow for modes with the additional frequencies that arise from the new pendulum motions of the thread.  Using data gathered from side-view images like that shown in Fig.~\ref{side_view}, amplitudes and phases of the constituent motions of the thread could be acquired and interpreted within the framework of the predictions of the amplitude equations.  Future work using this approach will require a more complete  investigation of the bifurcation structure of multifrequency amplitude equations with weakly broken $O(2)$ symmetry.

%Eqns.~\ref{eqn:amp1}~-~\ref{eqn:amp4}.  

\begin{acknowledgments}
This work was supported by the George Batchelor Laboratory, University
of Cambridge.  We acknowledge useful discussions with Sunny
Chiu-Webster, Sue Colwell,
%
% I talked to Sue about the amp eqns, and she very kindly worked
% through and checked some of my algebra - JHPD.
%
and Stuart Dalziel.  We
are grateful to Lucas Goehring for the rheometry and densitometry of
the silicon oil. JHPD is supported by Newnham College, Cambridge and
the Royal Society.
\end{acknowledgments}

\appendix*
\section{Resonant Hopf bifurcation with weakly broken $O(2)$ symmetry}
\label{appendix}

In this Appendix, we briefly sketch the derivation of the abstract amplitude
equations, Eqns.~\ref{eqn:amp1} and \ref{eqn:amp3}, describing the dynamics
for heights $H$ towards the lower end of the range investigated
experimentally. We emphasize that this discussion proceeds in a manner
independent of the fluid-mechanical details of the flow, and, although
it provides some support for our experimental results and simple
explanatory {\it ansatz}, it remains a modelling approach.

The system is taken to be a weak perturbation
of the axisymmetric case that corresponds to $U=0$. We first compute
the amplitude equations for the axisymmetric case and then discuss how
the weak symmetry-breaking provided by the belt admits additional
coupling terms.

%\subsection{Axisymmetric dynamics}  No labelled subsections in a
%short appendix. OK - JHPD.

In the axisymmetric case, the
symmetry group consists of planar rotations about the origin, and reflections in vertical planes containing the origin: this is the orthogonal group $O(2)$. 
The action of $O(2)$ on the plane $\mathbb{R}^2$
containing the belt is generated by the elements $\kappa$
(a reflection in the plane containing the belt) and
$\rho_\theta$ (rotation through angle $\theta$). Identifying $(x,y)
\in \mathbb{R}^2$ with $p=x + i y \in \mathbb{C}$ we have
%
%  I am not sure about notation like \in \mathbb{C}, which is pretty unusual in PRE.
% But a certain mathiness is probably appropriate in this appendix!
%
%
%  Here we have z the complex plane, when it was the perpendicular coordinate in the main paper
% Is this a problem? I've now used p=x+iy to avoid conflict. JHPD.
%
\begin{eqnarray}
\kappa (p)      = \bar{p}, \qquad \qquad
\rho_\theta (p) = \e^{i \theta} p. \nonumber
\end{eqnarray}
In addition, the
system has no preferred origin in time, and hence the time-periodic
solutions we investigate have no preferred phase. This additional
symmetry $\tau_\phi: t \rightarrow t+\phi/\omega$
generates a group $S^1$ corresponding to phase shifts of the
amplitudes. We suppose that the coiling motion can be described as a
sum of four oscillatory amplitudes for a complex position coordinate, as in Eqn.~\ref{4amps}.
%\ba
%x(t) + i y(t) & = & A_+(t) \e^{i \omega t} + A_-(t) \e^{-i \omega
 % t} \nn \\ 
%& & + B_+(t) \e^{2i \omega t} + B_-(t) \e^{-2i \omega t}. \nn
%\ea
Let $\mathbf{A}=(A_+,A_-,B_+,B_-)$ denote the vector of amplitudes. 
The action of the full symmetry group $O(2) \times S^1$ on $\mathbf{A}$
is inherited from the action on $\mathbb{R}^2$
of the generators $(\kappa,\rho_\theta)$,
and the phase-shift symmetry $\tau_\phi$:
\begin{eqnarray}
\kappa(A_+,A_-,B_+,B_-) & = & (\bar{A}_-, \bar{A}_+,\bar{B}_-,\bar{B}_+), \label{eqn:equiv1} \\
\rho_\theta (A_+,A_-,B_+,B_-) & = & \e^{i \theta} (A_+, A_-,B_+,B_-), \\
%
%
%  SWM:  Where did \tau_\phi come from here?  It seems to have appeared out of thin air ...
%  shouldn't it be defined with \kappa and \rho_\theta above??
%
%
\tau_\phi (A_+, A_-,B_+,B_-) & = & (\e^{i \phi} A_+, \e^{-i \phi} A_-,
\e^{2i \phi} B_+,\e^{-2i \phi} B_-). \nn \\ & & \label{eqn:equiv3}
\ea
The required amplitude equations
\begin{eqnarray}
\mathbf{\dot{A}} & = & \mathbf{F}(\mathbf{A}), \label{eqn:allamps}
\end{eqnarray}
that are symmetric under Eqns.~\ref{eqn:equiv1} -
\ref{eqn:equiv3} obey the equivariance condition
%\begin{eqnarray}
%g \mathbf{F} (\mathbf{A}) & = & \mathbf{F} ( g \mathbf{A}), \nn
$\gamma \mathbf{F} (\mathbf{A}) = \mathbf{F} ( \gamma \mathbf{A})$, % \nn
%\end{eqnarray}
for all group elements $\gamma \in O(2) \times S^1$. The equivariance
condition is a mathematical statement of the intuitive notion that
whenever $\mathbf{A}(t)$ is a solution of the amplitude equations, we
require $\gamma \mathbf{A}(t)$ also to be a solution trajectory.
We can now determine the form of the most general amplitude
equations by computing invariant and
equivariant polynomials. General methods for this exist~\cite{GSS88}
%
%  supply a weighty reference for this cumbersome procedure. Done - JHPD.
%
but are too cumbersome to apply here, particularly since we intend
only to compute terms at the first few orders.
%
% This next bit has been rewritten to reflect JRL's earlier comments - JHPD.
%
As usual in problems of this type, the terms
$|A_+|^2$, $|A_-|^2$, $|B_+|^2$ and $|B_-|^2$ (and sums and products of
these) are invariant. To find other invariants we 
adopt the ``brute force" approach of
looking for an invariant term $A_+^a A_-^b B_+^c B_-^d$ with integer
powers, using the convention that a negative power denotes
a positive power of the complex conjugate variable. 
Invariance under $\rho_\theta$ and $\tau_\phi$
implies $a+b+c+d=0$ and $a-b+2(c-d)=0$. We define the order of the
invariant term to be $|a|+|b|+|c|+|d|$. The invariants of
order 2 are $|A_+|^2$, $|A_-|^2$, $|B_+|^2$ and $|B_-|^2$. It is
easily seen that the only
invariant of order 4 that is not a product of order 2 invariants
is $\bar{A}_+ \bar{A}_- B_+ B_-$. It follows that the terms up to cubic
order in the first component
$\dot{A}_1=F_1(A_+,A_-,B_+,B_-)$ of~(\ref{eqn:allamps}) will be
\ba
A_+,\phantom{|B_+|^2}          \quad  A_+|A_+|^2, \quad A_+ |A_-|^2, \nn \\
A_+|B_+|^2, \quad  A_+|B_-|^2, \quad \bar{A}_- B_+ B_-.              \nn
\ea
By now applying the reflection symmetry $\kappa$, we may deduce the
second amplitude equation $F_2(A_+,A_-,B_+,B_-)$.
Note that $\kappa$ implies that the coefficients
in $F_2$ are the complex conjugates of those in $F_1$.

Similarly we deduce the order 2 and order 4 invariants for $F_3$ and
then use $\kappa$ to compute $F_4$ from $F_3$. The resulting amplitude
equations, Eqns.~\ref{eqn:amp1} and \ref{eqn:amp3}  with $\epsilon_b \equiv 0$, 
describe the axisymmetric interaction of the $\omega$ and $2\omega$
modes.

%It is interesting to note that, at least to cubic order, only a
%single combination of the phases of the amplitudes is important:
%$\arg{A_+}+\arg{B_+} - \arg{A_-}-\arg{B_-}$. This can be seen by
%writing the amplitudes in polar coordinates and taking the
%imaginary part of $F_1$.
%
% what does this mean physically?  That pure coiling always coils and does not meander??
% It's easier to leave this last sentence out; it isn't used later at all - JHPD.
%

The introduction of a (slowly-moving) horizontal belt changes the
symmetry of the problem in two ways. The first of these is clear: it
breaks the rotation symmetry $\rho_\theta$. 
To account for this effect we
compute low-order terms that are not symmetric under
$\rho_\theta$ but which do respect the remaining symmetry $\tau_\phi$. 
%
%I DECIDED THIS BIT COULD BE OMITTED IN THE INTERESTS OF BREVITY. JHPD.
%There are eight invariant terms $A_+^a A_-^b B_+^c
%B_-^d$ of order 2 and 3 that satisfy $a-b+2(c-d)=0$ but not
%$a+b+c+d=0$. These contribute new linear and quadratic terms to the
%amplitude equations. At higher orders the number of additional terms
%quickly becomes very large. In any case, we expect that the salient
%effects of the weak symmetry-breaking are controlled by terms at the
%lowest orders.
%
The eight new invariants are: 
\ba
A_+ A_-, \quad B_+ B_-, \quad \bar{A}_+ A_- B_+, \quad A_-^2 B_+, \nn \\
A_+^2 \bar{B}_+, \quad A_+\bar{A}_- B_-, \quad A_-^2 \bar{B}_-, \quad
A_+^2 B_-. \nn 
\ea
The new quadratic invariants give rise to the new linear terms
in Eqns.~\ref{eqn:amp1} and \ref{eqn:amp3} 
and so split the symmetric
Hopf bifurcation at $\epsilon_1=0$ into two generic Hopf bifurcations
at $\epsilon_1 \pm \epsilon_b {\rm Re}(e_{1})=0$. 
The new cubic invariants generate the new
quadratic terms.  The amplitude equations for the four amplitudes are then given by Eqn.~\ref{eqn:amp1} and \ref{eqn:amp3} along with
%
%
%  2 of the 4 full amp equations reproduced here  post-referee
%
%
%\begin{eqnarray}
%\dot{A}_+ & = & A_+ \left(\epsilon_1 + i \omega_1
%+ a_1 |A_+|^2 + a_2 |A_-|^2 \right. \nn \\
%& & \left. + b_1 |B_+|^2 + b_2 |B_-|^2 \right) + s_1 \bar{A}_- B_+ B_- \nn \\
%& & + \epsilon_b ( e_1  \bar{A}_- + p_1 \bar{A}_+ B_+ \nn \\
%& & + q_1 \bar{A}_+ \bar{B}_- + p_3 A_- B_+ + q_3 A_- \bar{B}_-), \label{eqn:appendixamp1} 
%  \end{eqnarray}
 %
\begin{eqnarray}
\dot{A}_- & = & A_- \left(\epsilon_1 - i \omega_1
+ \bar{a}_1 |A_-|^2 + \bar{a}_2 |A_+|^2 \right. \nn \\
& & \left. + \bar{b}_1 |B_-|^2 + \bar{b}_2 |B_+|^2 \right) + \bar{s}_1
\bar{A}_+ B_+ B_- \nn \\
& & +\epsilon_b ( \bar{e}_1 \bar{A}_+ + \bar{p}_1 \bar{A}_- B_-  \nn \\
& & + \bar{q}_1 \bar{A}_- \bar{B}_+ + \bar{p}_3 B_- A_+ + \bar{q}_3 A_+ \bar{B}_+), \label{eqn:appendixamp2}
   \end{eqnarray}
%
%\begin{eqnarray}
%\dot{B}_+ & = & B_+ \left(\epsilon_2 + i \omega_2
%+ c_1 |B_+|^2 + c_2 |B_-|^2 \right. \nn \\
%& & \left. + d_1 |A_+|^2 + d_2 |A_-|^2 \right) + s_2 \bar{B}_- A_+ A_- \nn \\
%& & + \epsilon_b ( e_2  \bar{B}_- + p_2 \bar{A}_-^2 + q_2 A_+^2 + p_4 A_+ \bar{A}_-),
%\label{eqn:appendixamp3} 
%  \end{eqnarray}
%
\begin{eqnarray}
\dot{B}_- & = & B_- \left(\epsilon_2 - i \omega_2
+ \bar{c}_1 |B_-|^2 + \bar{c}_2 |B_+|^2 \right. \nn \\
& & \left. + \bar{d}_1 |A_-|^2 + \bar{d}_2 |A_+|^2 \right) + \bar{s}_2 \bar{B}_+ A_+ A_- \nonumber \\
& & +\epsilon_b ( \bar{e}_2 \bar{B}_+ + \bar{p}_2 \bar{A}_+^2 +
  \bar{q}_2 A_-^2 + \bar{p}_4 \bar{A}_+ A_-). \label{eqn:appendixamp4}
\end{eqnarray}

~~

%Secondly, and more abstractly, the moving belt 
%
%  Not clear why the belt would do this: change words here post-referee
%

Some features of the phase relationship between the $\omega$ and $2\omega$ modes
suggest the introduction of an additional symmetry, denoted $h$: rotating the apparatus by
$180^\circ$ around the vertical $z$-axis and reversing the direction
of travel of the belt ({\it i.e.} reversing the direction of time) leaves
the basic state $p=0$ unchanged. For the bifurcation problem
constructed in this Appendix, $h$ is an independent symmetry that
introduces a new constraint. But $h$ is clearly not an exact symmetry of the
original fluid-mechanical problem where the fluid in the viscous
thread falls under gravity, and the initial equilibrium state is a
catenary rather than an exactly vertical thread.
%, at least at low belt speeds where the
%amount of horizontal stretching in the fluid thread is small (see
%Fig.\ref{state_photos}). \marginpar{JRL}[[I don't believe this is true.]]
%Substantially rewritten here. JHPD.
So, at best,  the effect of $h$ is felt in some approximate way by
amplitude equations describing the original physical system.

The new symmetry $h(x,y,t)=-(x,y,t)$ acts on the mode amplitudes as
\ba
h(A_+,A_-,B_+.B_-) & = & -(A_-,A_+,B_-,B_+). \label{eqn:equiv4}
\ea
So the composite operation $h \circ \kappa$ sends each amplitude to the
negative of its complex conjugate; equivariance of Eqn.~\ref{eqn:allamps} under
$h\circ \kappa$ implies that odd order terms would have real coefficients
and even order terms would have purely imaginary coefficients. More
realistically, we might hope that the coefficients have non-zero but
small (compared to their modulus) imaginary
or real parts at odd and even orders respectively.

%Hence the symmetry-breaking plays a key role in stabilising
%the relative phases between the modes and thus determining the
%nature of the patterns in which the $\omega$ and $2\omega$ modes
%interact strongly.

%\bigskip

%One final remark: 
The analysis of Eqns.~\ref{eqn:amp1}, \ref{eqn:amp3}, \ref{eqn:appendixamp2} and \ref{eqn:appendixamp4}
is simplified by the existence of invariant subspaces
for the dynamics - technically, the fixed-point subspaces of group
elements. The most relevant of these for this paper consists of those
points left unchanged by the combined operation $\kappa \circ
\tau_\pi$ (reflection in the plane of the belt followed by a
time-shift of half an oscillation period):
it is therefore useful to define the invariant subspace 
\ba
\S \equiv \Fix(\kappa \circ \tau_\pi) & = & (A_+,-\bar{A}_+,B_+,\bar{B}_+), \nn
%
%  Not sure if PRE will like this Fix notation ....
%  But is in the appendix - I've softened the notation in the main
%  part of the paper. JHPD.
%
\ea
which contains both the meandering and figure eight states.
%
%
%  Is figure 8 in here? Yes. JHPD.
%
%

% NOT NEEDED REALLY- OMIT. JHPD.
%These equations show that in general a small amplitude transverse
%meandering state ($A_+ \neq 0$) with frequency $\omega$ generates a 
%component $B_+$ at frequency
%$2\omega$ (at even smaller amplitude), i.e. the
%small figure-eight-like wobble shown in
%figure~\ref{side_view}(a). 

\end{document}